\documentclass[
aps,
pra,
reprint,
superscriptaddress,
amssymb,
amsmath,
amsfonts,
floatfix,
nobalancelastpage,
groupedaddress
]{revtex4-2}

\usepackage{mathtools}
\usepackage{graphicx}
\usepackage{graphics}
\usepackage{mathrsfs}

\usepackage{bm}
\usepackage{physics}
\usepackage{url}
\usepackage{hyperref}
\hypersetup{
	bookmarksnumbered,		
	unicode,			
	colorlinks,			
	citecolor=[rgb]{.9,0,.5},	
	urlcolor=[rgb]{0,0,1},		
	linkcolor=[rgb]{0,.7,0}		
	}

\usepackage[stable]{footmisc}
\usepackage{tensor}
\usepackage[super]{nth}
\usepackage{color}

\newcommand{\inv}{^{-1}}
\newcommand{\qqor}{\,\textnormal{or}\,}
\newcommand{\eff}{{\textnormal{eff}}}
\newcommand{\sgn}{\textnormal{sgn}}

\newcommand{\bdg}{{\textnormal{BdG}}}

\DeclareMathOperator{\diag}{diag}

\newcommand{\etal}{\textit{et al.\ }}
\newcommand{\tmax}{{\textnormal{max}}}
\newcommand{\pf}{\textnormal{Pf}}
\newcommand{\tint}{\textnormal{int}}
\newcommand{\krin}[1]{\langle\!\langle #1 \rangle\!\rangle}

\newcommand{\tgp}{\textnormal{GP}}
\newcommand{\hc}{\textnormal{h.c.}}

\newcommand{\bog}{\textnormal{Bog.}}

\begin{document}

\title{Pseudo-time-reversal-symmetry-protected topological Bogoliubov excitations of Bose-Einstein condensates in optical lattices}
\author{Junsen Wang}
\email{jsw@mail.ustc.edu.cn}
\author{Wei Zheng}
\author{Youjin Deng}
\affiliation{Hefei National Laboratory for Physical Sciences at Microscale and Department of Modern Physics,
University of Science and Technology of China, Hefei, Anhui 230026, China}
\affiliation{CAS Center for Excellence and Synergetic Innovation Center in Quantum Information and Quantum Physics,
University of Science and Technology of China, Hefei, Anhui 230026, China}
\date{\today}
\begin{abstract}
	Bogoliubov excitations of Bose-Einstein condensates in optical lattices may possess band topology in analogous to topological insulators in class AII of fermions. Using the language of the Krein-space theory, this topological property is shown to be protected by a pseudo-time-reversal symmetry that is pseudo-antiunitary and squares to $-1$, with the associated bulk topological invariant also being a $\mathbb Z_2$ index. We construct three equivalent expressions for it, relating to the Pfaffian, the pseudo-time-reversal polarization, and most practically, the Wannier center flow, all adopted from the fermionic case, defined here with respect to the pseudo inner product. In the presence of an additional pseudo-unitary and pseudo-Hermitian inversion symmetry, a simpler expression is derived. We then study two toy models feasible on cold atom platforms to numerically confirm the bulk-boundary correspondence. The Krein-space approach developed in this work is a universal formalism to study all kinds of symmetry-protected topological bosonic Bogoliubov bands.
\end{abstract}
\maketitle


\section{Introduction}
Topological band theory \cite{Qi2011,Hasan2010,Chiu2016} is originally developed to characterize nontrivial topology of electronic bands in solids. One milestone work in the early years along this direction is the tenfold-way classification of topological insulators and topological superconductors according to three non-spatial symmetries: time-reversal, particle-hole and chiral symmetry \cite{Altland1997,Zirnbauer2018}. Soon after, it is found that with additional spatial symmetries, topological invariants may have simplified expressions \cite{Fu2007}, or even the topological classification is enriched \cite{Fu2011}. Recently, topological properties of dynamical \cite{Chang2018,Yang2018,Gong2018,Qiu2018} and open quantum-mechanical systems \cite{Shen2018,Kawabata2018,Yao2018,Kawabata2019} are also studied extensively.

Since F. D. M. Haldane pointed out that topological band theory is not tied to fermions, but essentially wave effects \cite{Raghu2008}, there are many works focusing on topological phenomena in other physical systems, such as magnonic crystals \cite{Shindou2013,Chisnell2015,Kondo2019}, photonic crystals \cite{Wang2008,Rechtsman2013,Peano2016}, phononic crystall \cite{Fleury2014,SafaviNaeini2014,Peano2015} and even coupled pendula \cite{Suesstrunk2015}. Here we will focus on topology of collective modes in the weakly interacting ultracold bosonic atoms loaded in an optical lattice. It has been known that wave functions of the excited modes above Bose-Einstein condensates (BEC) can exhibit a topological structure \cite{Furukawa2015,Xu2016,Liberto2016}. However, all of these previous works focusing on systems \textit{breaking} time-reversal symmetry, which have a nonvanishing first Chern number in two dimensions that is in one-to-one correspondence with the number of chiral edge states dictated by the bulk-boundary correspondence. This type of excitation band topology is in analogous to the Chern insulators in class A. 

It is well-known that there is a topological phase protected by the odd time-reversal symmetry in two and three dimensions due to Kane and Mele \cite{Kane2005}, namely the topological insulators in class AII. This topological phase possesses a pair of helical edge states propagating in opposite directions, whose presence or absence is characterized by a $\mathbb Z_2$ index, which has many equivalent definitions, e.g., relating to the Pfaffian \cite{Kane2005}, the time reversal polarization \cite{Fu2006}, and the Wannier center flow \cite{Yu2011}. The last one is of most practical use, since it does not involve any gauge-fixing problems that occur in previous two definitions. One natural question to ask is whether similar topological structure exist in the excitation spectrum of a bosonic superfluid; if so, what type of symmetry protects them, how to define the associated bulk topological invariant, and whether the bulk-boundary correspondence still hold or not.

In this work, we show that a AII-class-like topological structure indeed exists in the Bogoliubov excitations of a BEC in an optical lattice, which is protected by a pseudo-time-reversal symmetry that is pseudo-antiunitary and squares to minus one. The corresponding bulk topological index is also a $\mathbb Z_2$ number, and three equivalent definitions used in the fermionic case all have counterparts here.

To address the problem in a systematical way that can be generalized easily for various kinds of symmetry-protected topological bosonic Bogoliubov bands, we review the problem of quadratic boson in Sec.~\ref{sec1}, where we also introduce the Krein-space theory to reformulate the problem in a way that has the closest connection to its fermionic counterpart. In Sec.~\ref{sec2}, we show that a pseudo-time-reversal symmetry, which is pseudo-antiunitary and squares to minus one, guarantees the bosonic Kramers' pair, which in turn protects a AII-class-like topological excitation spectrum. We then go on discussing three equivalent characterizations of the associated bulk $\mathbb Z_2$ topological invariant. Moreover, with an additional inversion symmetry, a simpler formula for it is derived. In Sec.~\ref{sec3}, two toy models feasible in cold atom experiments are studied, and the bulk-boundary correspondence is confirmed numerically. In Sec.~\ref{sec5} we conclude the paper and give a discussion.


\section{Preliminary}\label{sec1}


\subsection{Quadratic boson}\label{sec11}
Consider loading ultracold bosonic atoms in an optical lattice. In the weakly interacting region, atoms condense in the single-particle ground state. By employing the standard Bogoliubov-de Gennes approximation, the excitation of the condensate can be well described by a bosonic quadratic Hamiltonian, whose most general form in real space reads,
\begin{equation}\label{fb}
\begin{split}
	H =&\sum_{\vb r,\vb r'}\sum_{\alpha\beta}a_{\vb r \alpha}^\dagger A_{\vb r\alpha,\vb r'\beta}  a_{\vb r' \beta}\\
&+ \frac{1}{2}\sum_{\vb r,\vb r'}\sum_{\alpha\beta}\pqty{a^\dagger_{\vb r \alpha}B_{\vb r \alpha,\vb r' \beta}a^\dagger_{\vb r' \beta} +a_{\vb r \beta} B^*_{\vb r \alpha,\vb r' \beta}a_{\vb r' \alpha}},
\end{split}
\end{equation}
where the bosonic creation (annihilation) operators $a_{\vb r \alpha}^{(\dagger)}$, labeled by an external index $\vb r$ (assuming total $M$ unit cells) and an internal index $\alpha\in {1,2, \dots ,N}$, satisfy the canonical commutation relations (CCRs) $[a_{\vb r \alpha},a^\dagger_{\vb r' \beta}]=\delta_{\vb r\vb r'}\delta_{\alpha\beta}$ and $[a_{\vb r \alpha},a_{\vb r' \beta}]=[a^\dagger_{\vb r \alpha},a^\dagger_{\vb r' \beta}]=0$. Using a single index, one can write $a$ and $a^\dagger$ as $MN$-dimensional arrays, and $A$ and $B$ as $MN \times MN$ matrices. We then have $A=A^\dagger$ due to Hermiticity of $H$, and $B^T=B$ due to Bose statistics. Further introducing the Nambu spinor $\alpha=\begin{pmatrix}
	a \\
	a^\dagger
\end{pmatrix}$ and $\alpha^\dagger=\begin{pmatrix}
	a^\dagger & a
\end{pmatrix}$, we can write Eq.~\eqref{fb} as
\begin{equation}\label{rsbdg}
	H=\frac{1}{2}\alpha^\dagger H_\bdg \alpha ,\quad H_\bdg =\begin{pmatrix}
		A & B\\
		B^* & A^T
	\end{pmatrix},
\end{equation}
where a constant term $\trace A$ is omitted and the block matrix $H_\bdg$ is called the Bogoliubov-de Gennes (BdG) Hamiltonian. Because of the particle-hole constraint of Nambu spinor $\alpha=\Sigma_1 (\alpha^{\dagger})^T$, where throughout this paper we denote $\Sigma_i = \sigma_i \otimes I_{MN}$, with $\sigma_i$ ($i=1,2,3$) being the standard Pauli matrices and $I_{MN}$ being the $MN \times MN$ identity matrix (in momentum space it becomes $I_N$), the BdG Hamiltonian enjoys a particle-hole ``symmetry" (PHS), $\mathcal C H_\bdg \mathcal C\inv = H_\bdg$, where $\mathcal C=\Sigma_1 K$ and $K$ is the complex conjugation. In the presence of translation symmetry, i.e., $A_{\vb r \alpha,\vb r'\beta}=A_{\vb r-\vb r', \alpha \beta}$ and similarly for $B$, Eq.~\eqref{rsbdg} can be written in momentum space as
\begin{equation}\label{fbbdg}
	H=\frac{1}{2}\sum_{\vb k}\alpha_{\vb k}^\dagger H_\bdg(\vb k) \alpha_{\vb k},\ H_\bdg(\vb k)=\begin{pmatrix}
		A_{\vb k} & B_{\vb k}\\
		B_{-\vb k}^* & A_{-\vb k}^T
	\end{pmatrix}
\end{equation}
where the $2N$-dimensional arrays are defined by
\begin{equation}
	\begin{split}
		\alpha_{\vb k}^\dagger &= \begin{pmatrix}
		a^\dagger_{\vb k1} &  \dots  &a^\dagger_{\vb kN} & a_{-\vb k1} &  \dots  &a_{-\vb kN}
		\end{pmatrix},\\
		\alpha_{\vb k} &= \begin{pmatrix}
		a_{\vb k1} &  \dots  &a_{\vb kN} & a^\dagger_{-\vb k1} &  \dots  &a^\dagger_{-\vb kN}
		\end{pmatrix}^T.
	\end{split}
\end{equation}
The $N\times N$ Hermitian matrix $A_{\vb k}$ has entries $[A_{\vb k}]_{\alpha\beta}=\sum_{\vb r}A_{\vb r-\vb r',\alpha\beta}e^{i\vb k\cdot\vb r}$ and similarly for the symmetric matrix $B_{\vb k}$. The PHS now reads $\mathcal C H_\bdg(\vb k) \mathcal C\inv =H_\bdg(-\vb k)$. In this representation, CCRs take a compact form
\begin{equation}
	[\alpha_{\vb ka},\alpha_{\vb k'b}^\dagger]=[\Sigma_3]_{ab} \delta_{\vb k,\vb k'}.\label{cr}
\end{equation}

Eq.~\eqref{fbbdg} [and similarly for Eq.~\eqref{rsbdg}] is solved by a Bogoliubov transformation
\begin{subequations}\label{btdef}
\begin{align}
	\beta_{\vb k} &= T_{\vb k}\alpha_{\vb k}, \label{btdef1} \\
	\beta^\dagger_{\vb k} &= \beta^T_{-\vb k} \Sigma_1=\alpha^\dagger_{\vb k} \Sigma_1 T^T_{-\vb k} \Sigma_1, \label{btdef2}
\end{align}
\end{subequations}
where the second equality in Eq.~\eqref{btdef2} results from Eq.~\eqref{btdef1}. The transformation is assumed (i) To be canonical, i.e., the CCRs are preserved, (double indices imply summation)
\begin{align}
  	[\Sigma_3]_{ab} &=[\beta_{\vb k a},\beta_{\vb k b}^\dagger] \nonumber \\
  	&= [T_{\vb k}]_{aa'}[\alpha_{\vb ka'},\alpha^\dagger_{\vb kb'}][\Sigma_1T^T_{-\vb k}\Sigma_1]_{b'b}\nonumber\\
  	&= [T_{\vb k} \Sigma_3 \Sigma_1 T_{-\vb k}^T \Sigma_1]_{ab} \nonumber\\
  	\Rightarrow i\Sigma_2 &= T_{\vb k} i\Sigma_2 T_{-\vb k}^T,\label{ty}
\end{align}
Note the real space version of Eq.~\eqref{ty} is $i\Sigma_2 = T i\Sigma_2 T^T$, where $\Sigma_2$ and $T$ are $MN\times MN$ matrices in this case. Hence the Bogoliubov transformations in real space form a complex symplectic group $\mathrm{Sp}(2NM,\mathbb C)$ \footnote{See, e.g., \url{https://groupprops.subwiki.org/wiki/Symplectic_group} for a mathematical definition of symplectic group.}.
(ii) To be unitary, i.e.,
\begin{align}
	(\beta_{\vb ka})^\dagger &= \beta^\dagger_{\vb ka}\nonumber\\
	\alpha_{\vb ka'}^\dagger [T_{\vb k}^\dagger]_{a'a} &= \alpha^\dagger_{\vb ka'} [\Sigma_1 T^T_{-\vb k} \Sigma_1]_{a'a}\nonumber\\
	\Rightarrow T_{\vb k} &= \Sigma_1 T^*_{-\vb k} \Sigma_1. \label{tx}
\end{align}
Combining Eq.~\eqref{tx} and Eq.~\eqref{ty}, we further obtain
\begin{equation}
	T_{\vb k}\Sigma_3T^\dagger_{\vb k} \Sigma_3 =\Sigma_3T^\dagger_{\vb k} \Sigma_3T_{\vb k}=I_{2N} .\label{tz}
\end{equation}
Matrices satisfying Eq.~\eqref{tz} are called para-unitary by Colpa \cite{Colpa1978}. (iii) The transformed Hamiltonian $H=\frac{1}{2}\beta_{\vb k}^\dagger \Sigma_3 T_{\vb k} \Sigma_3 H_\bdg(\vb k) T_{\vb k}\inv \beta_{\vb k}$ is diagonal, i.e.,
\begin{equation}
	\begin{split}
		&T_{\vb k}\Sigma_3 H_\bdg(\vb k) T_{\vb k}\inv\\
		& =\diag(
		E_1(\vb k),\dots, E_N(\vb k),-E_1(-\vb k),\dots,-E_N(-\vb k)),\label{tsz}
	\end{split}
\end{equation}
with $E_1(\vb k)\leq \dots \leq E_{N}(\vb k)$. In this paper, we assume $H_\bdg$ is positive semidefinite, so that all eigenvalues are real \cite{Blaizot1986}. Eq.~\eqref{tsz} indicates that one has to solve an eigenproblem of a generally \textit{non-Hermitian} matrix [here only $H_{\bdg}(\vb k)$ itself is Hermitian by definition]
\begin{equation}
	H^\eff_{\vb k} = \Sigma_3H_\bdg(\vb k),\label{defhk}
\end{equation}
with a PHS,
\begin{equation}
	\mathcal C H^\eff_{\vb k} \mathcal C\inv = -H_{-\vb k}^\eff,\nonumber
\end{equation}
which guarantees that eigenvalues always come in pairs as shown in Eq.~\eqref{tsz}: for each eigenstate $\ket{u^+_n(\vb k)}$ with nonnegative eigenvalue $E_n(\vb k)$, called the particle excitation, we have another eigenstate, $\ket{u^{-}_{n}(\vb k)}\coloneqq \mathcal C\ket{u^{+}_n(-\vb k)}=\Sigma_1 \ket{u^{+}_n(-\vb k)^*}$ called the hole excitation, with nonpositive eigenvalue $-E_n(-\vb k)$.

One may alternatively arrive at the effective Hamiltonian by examining the Heisenberg equations of motion for the operator $\alpha_{\vb k}(t)$ \cite{Kim1988,Zhang1990,Richaud2017},
\begin{equation*}
	i\dv{t}\alpha_{\vb ka}(t)=[\alpha_{\vb ka}(t),H]=[\Sigma_3 H_\bdg(\vb k)]_{aa'}\alpha_{\vb ka'}(t)
\end{equation*}
where Eq.~\eqref{cr} is used in the second equality. Hence the dynamics of the system is indeed generated by the non-Hermitian matrix Eq.~\eqref{defhk}.


\subsection{Krein space formalism}\label{ksf}
Here we review the basics of the Krein-space theory and reformulate the problem of quadratic boson using this language \cite{Peano2018,Bender2019,Lein2019}.

A Krein space $(\mathcal K,J)$ is a Hilbert space $\mathcal K$ with a \textit{fundamental symmetry} $J$ which is a linear operator satisfying $J^2=1$ and $J=J^\dagger$. Equivalently, operator $J$ is a fundamental symmetry if
\begin{equation}
	J^2=1, \qand \expval{J\phi,J\psi}=\expval{\phi,\psi},\ \forall \phi,\psi\in \mathcal K,\nonumber
\end{equation}
where $\expval{\cdot,\cdot}$ is the usual inner product in the Hilbert space $\mathcal K$. A Krein space becomes real if there is a real structure and a real unitary $Q$ which squares to $\pm 1$ and (anti)commutes with $J$ \cite{SchulzBaldes2016}.

We define the \textit{pseudo inner product} as
\begin{equation}
	\krin{\phi,\psi} \coloneqq \expval{\phi,J \psi}. \label{defkip}
\end{equation}
It then follows that all familiar concepts defined w.r.t. the usual inner product have a pseudo-inner-product version. First of all, the \textit{pseudo-adjoint} is defined by $A^\sharp=J A^\dagger J$, which by definition satisfies $\krin{A^\sharp \phi,\psi}=\krin{\phi,A\psi}$. Then the \textit{pseudo-Hermitian} means $A^\sharp=A$, namely, Hermitian w.r.t. the pseudo inner product. \textit{Pseudo-unitary} means $A^\sharp=A\inv$, namely, its pseudo-adjoint equals its inverse. \textit{Pseudo-antiunitary} is antilinear w.r.t. the pseudo inner product, i.e., $\krin{A\phi,A\psi}=\krin{\phi,\psi}^*=\krin{\psi,\phi}$. The \textit{pseudo-orthogonal projector} is an operator that squares to itself and is pseudo-Hermitian, $\Pi^2=\Pi=\Pi^\sharp$, which implies that $\Pi$ and $\Pi^\dagger$ are related by a similarity transformation.

Unlike Hermiticity, pseudo-Hermiticity does not guarantee reality of the spectrum. Nevertheless, the \textit{Krein-spectral operator} $H$, defined by
\begin{equation}
	\tilde H=UHU\inv =UHU^\sharp=\tilde H^\sharp =\tilde H^\dagger,\nonumber
\end{equation}
has real spectrum. Operators that are non-negative w.r.t. the pseudo inner product are automatically Krein-spectral \cite{Peano2018,Colpa1978}.

For the bosonic BdG system studied in this paper, we have the real Krein space of kind $(1,-1)$ \cite{SchulzBaldes2016} by setting $J=\Sigma_3$ and $Q=\Sigma_1$. The effective Hamiltonian Eq.~\eqref{defhk} is pseudo-Hermitian with a real symmetry. The Bogoliubov transformation matrix $T_{\vb k}$ Eq.~\eqref{btdef} is pseudo-unitary with a real symmetry.

Eq.~\eqref{tz} can be rewritten using the pseudo inner product as,
\begin{equation}\label{orthogonal conditions}
	\begin{split}
		\krin{u^\pm_n(\vb k),u^\pm_m(\vb k)} &=\pm \delta_{mn},\\
		\krin{u^\pm_n(\vb k),u^\mp_m(\vb k)} &=0.
	\end{split}
\end{equation}
where $\ket{u^\pm_n(\vb k)}$ is the right-eigenstate of $H^\eff_{\vb k}$ with eigenvalue $\pm E_n(\pm \vb k)$. The pseudo-orthogonal projector, which is pseudo-Hermitian and generally non-Hermitian, then takes the form
\begin{equation}\label{projop}
	\Pi_{n,\vb k}=\pm \ketbra{u^\pm_n(\vb k)}{u^\pm_n(\vb k)}\Sigma_3.
\end{equation}
From here on, unless otherwise stated, we will focus on the particle space since the hole excitations are just copies of the former due to PHS. To prevent cluttering, the superscript ``+", indicating states in particle space, will be omitted.

We briefly mention how to deal with the so-called Nambu-Goldstone (NG) modes in the context of topological band theory before proceeding further. The NG mode is a gapless mode due to spontaneously breaking a continuous symmetry. There are several types of NG modes; they may or may not satisfy the orthonormal conditions Eq.~\eqref{orthogonal conditions} \cite{Takahashi2015,Watanabe2020}. Nevertheless, this type of modes can always be removed by adding an infinitesimal external field that explicitly breaks the corresponding symmetry. E.g., the gapless phonon modes can open an infinitesimal gap by shifting the chemical potential in the negative direction, $\mu\rightarrow \mu-0^+$. For simplicity, in this paper, we assume that the NG mode will have an infinitesimal gap. In fact, as will be discussed in the end of next section, when considering the bulk-boundary correspondence to determine the helical mid-gap edge states, one may completely avoid discussing the topology relating to the lowest particle bands and highest hole bands (also see Ref.~\cite{Furukawa2015}).
\section{$\mathbb Z_2$ invariant associated with pseudo-time-reversal symmetry}\label{sec2}
\subsection{Pseudo-time-reversal symmetry}
For a bosonic system, the conventional time-reversal symmetry squares to $+1$ \cite{Sakurai2014}. We define a pseudo-time-reversal (PTR) operator $\mathcal{T}=PK$ that squares to $-1$. Here $P$ is a $\vb k$-independent pseudo-unitary matrix. By definition, $\mathcal T$ is pseudo-antiunitary \footnote{The proof is similar to the case for the ordinary TRS. Consider
\begin{eqnarray*}
	\krin{\phi,\mathcal{T}\psi} &=& \phi^*_i (\Sigma_3P)_{ij} \psi^*_j\\
	&=& \psi_j^* (P^T\Sigma_3)_{ji}\phi^*_i\\
	&=& \krin{\psi,\Sigma_3 P^T\Sigma_3 K\phi},
\end{eqnarray*}
then replacing $\phi$ by $\mathcal{T}\phi$ and using pseudo-unitarity of $P$,
\begin{eqnarray*}
	\krin{\mathcal{T}\phi,\mathcal{T}\psi} &=& \krin{\psi,\Sigma_3 P^T \Sigma_3 KPK\phi}\\
	& = &\krin{\psi,\phi}.
\end{eqnarray*}
Hence $\mathcal T$ is indeed pseudo-antiunitary.}.

A bosonic BdG system is said to respect the pseudo-time-reversal symmetry (PTRS) if
\begin{equation}
	\mathcal{T} H^\eff_{\vb k} \mathcal{T}\inv=H_{-\vb k}^\eff.\label{PTRS}
\end{equation}
Implications of PTRS for bosons resembles that of odd TRS for fermions: For every eigenstate $\ket{u_n(\vb k)}$ of $H^\eff_{\vb k}$, due to Eq.~\eqref{PTRS}, $\mathcal{T}\ket{u_n(-\vb k)}$ is also an eigenstate with eigenvalue $E(-\vb k)$. At the pseudo-time-reversal-invariant momenta (PTRIM) $\vb{\Lambda}$, these two states have the same eigenenergy and are orthogonal w.r.t. the pseudo inner product \cite{Kondo2019}. The orthogonality can be seen by considering $\krin{\mathcal{T}\phi,\mathcal{T}\psi}=\krin{\psi,\phi}$, which, upon setting $\ket{\psi}=\ket{u_n(\vb{\Lambda})}$ and $\ket{\phi}=\mathcal{T} \ket{\psi}$, has to vanish separately on both sides due to $\mathcal{T}^{2}=-1$, leading to the orthogonality of bosonic Kramers' pair, $\ket{u_n(\vb{\Lambda})}$ and $\mathcal{T} \ket{u_n(\vb{\Lambda})}$.

\subsection{The Pfaffian approach}
In analogous to Kane and Mele's construction of $\mathbb Z_2$ invariant \cite{Kane2005}, consider the matrix of overlaps w.r.t. the pseudo inner product $\krin{u_n(\vb k),\mathcal T u_m(\vb k)}$, which is antisymmetric because $\mathcal T$ is pseudo-antiunitary and squares to $-1$. Assuming no other degeneracies, it is a $2\times 2$ matrix and can be written as
\begin{equation}\label{overlap}
	\krin{u_n(\vb k),\mathcal T u_m(\vb k)}=\epsilon_{nm} P(\vb k),
\end{equation}
with $P(\vb k)$ the Pfaffian of the matrix
\begin{equation}
	P(\vb k)=\pf[\krin{u_n(\vb k),\mathcal T u_m(\vb k)}].\label{defpf}
\end{equation}
Under a $\mathrm{U}(2)$ transformation $\ket{u_n(\vb k)} \rightarrow  R_{nm}(\vb k)\ket{u_m(\vb k)}$, the Pfaffian becomes $P(\vb k) \rightarrow  \det[R^*(\vb k)]P(\vb k)$ [cf. Eq.~\eqref{pdetrp}]. Thus $P(\vb k)$ is invariant under a $\mathrm{SU}(2)$ rotation but not $\mathrm{U}(1)$, since the latter induces an overall phase factor. Nevertheless, $\abs{P(\vb k)}$ is $\mathrm{U}(2)$ gauge invariant. At PTRIM $\vb \Lambda $, due to the existence of bosonic Kramers' pair, the off-diagonal element has unit modulus, namely $\abs{P(\vb \Lambda )}=1$. We further define the unitary sewing matrix $B(\vb k)$ that relates the PTR companion of an eigenstate at $\vb k$ to another eigenstate at $-\vb k$,
\begin{equation}
	\ket{u_m(-\vb k)}=B^*_{mn}(\vb k)\mathcal T \ket{u_n(\vb k)},\label{sewingm}
\end{equation}
which leads to $P(-\vb k)=\det[B(\vb k)] P^*(\vb k)$ [cf. Eq.~\eqref{pfatkmk}]. Thus whenever the Pfaffian vanishes at one momentum, so does the one at the opposite momentum with opposite ``vorticity." It then follows that the number of pairs of zeros of Pfaffian is a $\mathbb Z_2$ invariant in the presence of PTRS, due to the same reason as in the fermionic case \cite{Kane2005}. We hence conclude that the
winding of the phase of $P(\vb k)$ around a loop enclosing \textit{half} the first Brillouin zone (1BZ),
\begin{equation}
	I= \frac{1}{2\pi i}\oint_C \dd{\vb k}\cdot\nabla_{\vb k}\log[P(\vb k)],\nonumber
\end{equation}
is a $\mathbb Z_2$ invariant associated with PTRS for the bosonic BdG systems. It can be seen easily [cf. Eq.~\eqref{holepfaffian}] that the Pfaffian in the particle bands and their hole companion has the same number of zeros, i.e., $I_\textnormal{particle}=I_\textnormal{hole}$.
\subsection{Pseudo-time-reversal polarization}
One can also define a pseudo-time-reversal polarization to characterize this $\mathbb Z_2$ invariant in analogous to Fu and Kane \cite{Fu2006}, which is also a straightforward generalization of the symplectic ``charge" polarization constructed by Engelhardt and Brandes \cite{Engelhardt2015} for the bosonic BdG systems. Consider an effective one-dimensional (1D) system with $k_2$ (regarded as time $t$) fixed at $k_2=0\qqor \pi$ ($t=0\qqor T/2$), and set $k_1=k$. Assuming no other degeneracies, $N$ particle bands can be grouped into $N/2$ PTR pairs. The $\lambda $-th ($\lambda =1,2 \dots ,N/2$) pair for particle excitations are denoted by $\ket{u_{\lambda }^{(l)}(\vb k)}$, with $l=1,2$ labeling the two states of the pair. Due to PTRS, for each pair, the PTR companion of an eigenstate with $l=2$ at $\vb k$ equals the eigenstate with $l=1$ at $-\vb k$ up to a phase factor \cite{Fu2006},
\begin{subequations}\label{12x}
\begin{eqnarray}
	\ket{u_{\lambda }^{(1)}(-k)} &=& -e^{i\chi_{k,\lambda }}\mathcal T \ket{u_{\lambda }^{(2)}(k)},\\
	\ket{u_{\lambda }^{(2)}(-k)} &=& e^{i\chi_{-k,\lambda }}\mathcal T \ket{u_{\lambda }^{(1)}(k)}.
\end{eqnarray}
\end{subequations}
where the second equation results from the first one. We define the partial polarization for the $\lambda$-th pair by
\begin{equation}
	P_{\lambda }^{(l)}=\frac{1}{2\pi}\int_{-\pi}^\pi\dd{k} A_{\lambda }^{(l)}(k),\nonumber
\end{equation}
with the Berry connection \cite{Shindou2013}
\begin{equation}\label{berryconnectionpartial}
	A^{(l)}_{\lambda }(k) = i \krin{u_{\lambda }^{(l)}(k),\partial_k u_{\lambda }^{(l)}(k)}.
\end{equation}
The sum of two partial polarization is the symplectic generalization of ``charge" polarization \cite{Engelhardt2015}. Here we consider their difference,
\begin{equation}
	\tilde P_{\lambda }=P^{(1)}_\lambda -P^{(2)}_\lambda,\nonumber
\end{equation}
as the symplectic generalization of time-reversal polarization introduced by Fu and Kane \cite{Fu2006}, which satisfies \cite{supp}
\begin{equation}
	(-1)^{\tilde P_\lambda }=\frac{\sqrt{\det[B_\lambda (0)]}}{\pf[B_\lambda (0)]}\frac{\sqrt{\det[B_\lambda (\pi)]}}{\pf[B_\lambda (\pi)]},\label{trp}
\end{equation}
where $B_\lambda (k)=B(k,0)\qqor B(k,\pi)$ is the sewing matrix defined in Eq.~\eqref{sewingm} for the $\lambda $-th pair, and the sign ambiguity of the square root is fixed by requiring that $\sqrt{B_\lambda (k)}$ is continuous for $k\in [0,\pi]$. Follow the discussion in Ref.~\cite{Fu2006}, the \textit{change} in the PTR polarization during \textit{half} the cycle, which physically tracks the difference between positions of the pairs of Wannier states, defines a $\mathbb Z_2$ invariant (i.e., whether the Wannier states ``switch partners" or not),
\begin{equation}
	\Delta_\lambda =\tilde P_\lambda (T/2)-\tilde P_\lambda (0)\mod 2.\label{z2PTRp}
\end{equation}
Using Eq.~\eqref{trp}, we may equivalently write Eq.~\eqref{z2PTRp} as
\begin{equation}\label{z2dl}
	(-1)^{\Delta_\lambda }=\prod_{i=1}^4\frac{\sqrt{\det[B_\lambda (\Gamma_i)]}}{\pf[B_\lambda (\Gamma_i)]}.
\end{equation}
It is easily seen that $\Delta_\lambda $ is the same for particle bands and its hole companion, i.e., $\Delta_\lambda ^{\textnormal{particle}}=\Delta_\lambda ^{\textnormal{hole}}$ [cf. Eq.~\eqref{bhole}].

Incidentally, by enforcing the \textit{pseudo-time-reversal constraint} \cite{Fu2006},
\begin{equation*}
	\begin{split}
		\ket{u^{(1)}_\lambda (-k,-t)} &=\mathcal T\ket{u^{(2)}_\lambda (k,t)},\\
		\ket{u^{(2)}_\lambda (-k,-t)} &=-\mathcal T\ket{u^{(1)}_\lambda (k,t)},
	\end{split}
\end{equation*}
the $\mathbb Z_2$ invariant can also be interpreted as an obstruction. The resulting formula in terms of the Abelian Berry connection
\begin{equation}\label{berryconnection}
	\vb A_\lambda (\vb k)= \sum_{l=1,2} i\krin{u_\lambda ^{(l)}(\vb k),\nabla_{\vb k} u_\lambda ^{(l)}(\vb k)},
\end{equation}
and the Abelian Berry curvature
\begin{equation}
	F_{\lambda }(\vb k) =\sum_{l=1,2} [\nabla_{\vb k}\times \vb A_{\lambda }^{(l)}(\vb k)]_z,\nonumber
\end{equation}
has recently been proposed by Kondo \etal \cite{Kondo2019}, which is defined by
\begin{equation}\label{fukui}
	\begin{split}
		&\quad \tilde \Delta_\lambda =\\
	& \frac{1}{2\pi}\Bqty{\oint_{\partial \textnormal{HBZ}}\dd{\vb k}\cdot\vb A_\lambda (\vb k)-\int_{\textnormal{HBZ}}\dd[2]{k}F_\lambda (\vb k)}\mod 2,
	\end{split}
\end{equation}
where $\textnormal{HBZ}$ and $\partial\textnormal{HBZ}$ denotes half the 1BZ and its boundary that does not have any two points related by PTRS. The proof that $\tilde \Delta_\lambda =\Delta_\lambda $, and their equivalence to the Pfaffian approach can be obtained similarly as given in the Appendix of Ref.~\cite{Fu2006}.
\subsection{$\mathbb Z_2$ invariant as Wannier center flow}
All equivalent definitions of the $\mathbb Z_2$ invariant discussed so far suffer from the gauge-fixing problem. Here we generalize a practical Wilson loop approach proposed by Yu \etal \cite{Yu2011}. It is extremely useful for numerics because it does not require any gauge-ﬁxing condition.

Again consider the effective 1D system (with fixed $k_2$), we define the position operator $\hat X=\sum_{i\alpha}e^{i \boldsymbol{\delta}_1\cdot\vb r_i}\ketbra{i\alpha}{i\alpha}$ as usual \cite{Resta2000}, where $\ket{i\alpha}=\ket{i}\otimes\ket{\alpha}=a^\dagger_{\vb r_i \alpha}\ket{0}$ and $\boldsymbol{\delta}_1=\vb b_1/N_1$ with $\vb b_1$ the primitive reciprocal vector, and $N_1$ the number of unit cells for the effective 1D system. The Wannier states are given by the eigenstates of the position operator restricted in the occupied bands \cite{Kivelson1982}, $\hat X_P=\hat P\hat X \hat P$. Here the projection operator for the occupied subspace is [cf. Eq.~\eqref{projop}]
\begin{equation}
	\hat P = \sum_{n\leq n_\tmax,k_1} \ketbra{\psi_n(\vb k)}{\psi_n(\vb k)}\Sigma_3,\nonumber
\end{equation}
where $\ket{\psi_n(\vb k)}=\ket{\vb k}\otimes\ket{u_n(\vb k)}$ is the Bloch eigenstate, and $\ket{\vb k}=M^{-1/2} \sum_i e^{-i\vb k\cdot\vb r_i}\ket{i} $. Using
\begin{equation*}
	\krin{\psi_{n}(\vb k),\hat X \psi_{n'}(\vb k')} = \delta_{\vb k+\boldsymbol{\delta}_1,\vb k'}\krin{u_{n}(\vb k),u_{n'}(\vb k+\boldsymbol{\delta}_1)},
\end{equation*}
the projected position operator can be written as
\begin{equation*}
\begin{split}
	\hat X_P=\sum_{n,n'\leq n_\tmax}\sum_{k_1}\bigg[\krin{u_{n}(\vb k),u_{n'}(\vb k+\boldsymbol{\delta}_1)}\\
	\times \ketbra{\psi_{n}(\vb k)}{\psi_{n'}(\vb k+\boldsymbol{\delta}_1)}\Sigma_3\bigg]
\end{split}
\end{equation*}
We then raise $\hat X_P$ to the $N_1$-th power,
\begin{equation*}
	\hat X_P^{N_1}=\sum_{m,n\leq n_\tmax}\sum_{k_1}\bqty{W_{\vb k}}_{mn}\ketbra{\psi_m(\vb k)}{\psi_n(\vb k)}\Sigma_3
\end{equation*}
where the so-called \textit{Wilson loop operator} $W_{\vb k}$ is defined by
\begin{equation}
	W_{\vb k} =M^{(\vb k,\vb k+\boldsymbol{\delta}_1)} M^{(\vb k+\boldsymbol{\delta}_1,\vb k+2\boldsymbol{\delta}_1)}  \dots  M^{(\vb k+(N_1-1)\boldsymbol{\delta}_1,\vb k)},\label{wilsonloop}
\end{equation}
with $\bqty{M^{(\vb k,\vb k+\boldsymbol{\delta}_1)}}_{mn} =\krin{u_{m}(\vb k),u_n(\vb k+\boldsymbol{\delta}_1)}$. In the limit $\abs{\boldsymbol{\delta}}_1\rightarrow 0$, we have $\bqty{M^{(\vb k,\vb k+\boldsymbol{\delta}_1)}}_{mn}\rightarrow e^{-i[\mathcal A_1(\vb k)]_{mn} dk_1}$ with the non-Abelian $\mathrm{U}(n_m)$ gauge field defined by
\begin{equation}\label{nonaa}
	[\mathcal A_{1}(\vb k)]_{mn}=i \krin{u_m(\vb k),\partial_{k_1} u_n(\vb k)},
\end{equation}
and Eq.~\eqref{wilsonloop} becomes the $\mathrm{U}(n_m)$ Wilson loop, $W_{\vb k}=P\exp\bqty{\int_{-\pi}^\pi -i\mathcal A_1(\vb k)dk_1}$ \cite{peskin1995an}.

The eigenvalues of $W_{\vb k}$ are independent of $k_1$ and gauge invariant under a $\mathrm{U}(n_m)$ gauge transformation of $\ket{u_n(\vb k)}$ \cite{supp}. They are explicitly denoted by $w_n=\abs{w_n}e^{i\theta_n}$, for $n=1, \dots ,n_\tmax$ with $\theta_n\in (-\pi,\pi]$. The eigenvalues of $\hat X_P$, as the $N_1$-th roots of $w_n$, reads $w_{n,j}=\exp[i\theta_n/N_1+i 2\pi j/N_1 +(\log\abs{w_n})/N_1]$ for $j=1, \dots ,N_1$. Finally, the Wannier centers are identified with the phase of $w_{n,j}$,
\begin{equation}
	\expval{x}_{n,j}=\frac{N_1}{2\pi}\arg w_{n,j}=\expval{x}_n+j,\quad \expval{x}_n=\theta_n/2\pi\nonumber
\end{equation}
which is defined only up to a lattice translation.

Due to PTRS, eigenvalues of the Wilson loop operator at $k_2$ and $-k_2$ are the same; moreover, at PTRIM, eigenvalues are at least doubly degenerate \cite{supp}. Thus, starting from $k_2=-\pi$, each Wannier center pair will split and recombine at $k_2=0$; and for $k_2>0$, the behavior is just the mirror of the former w.r.t. the $k_2=0$ plane. Due to the same reason as in the fermionic case \cite{Yu2011}, the sum of winding numbers for all $n_\tmax/2$ Wannier center pair is a $\mathbb Z_2$ invariant. The equivalence of this definition to all previous ones can be obtained, although tedious, similarly as given in the Appendix of Ref.~\cite{Yu2011}.

\subsection{Simplifications from inversion symmetry}
Lastly, we show that with an additional inversion symmetry (IS), the $\mathbb Z_2$ invariant Eq.~\eqref{z2dl} takes a simple form,
\begin{equation}\label{isz2}
	(-1)^{\Delta_\lambda }=\prod_i \xi_\lambda (\vb \Lambda _i),
\end{equation}
where $\xi_\lambda (\vb \Lambda _i)$ is the parity eigenvalue of one of the $\lambda $-th pair of bands at the PTRIM $\vb \Lambda _i$. In analogous to the fermionic case discussed by Fu and Kane \cite{Fu2007}, we explicitly construct a globally continuous transverse gauge where $\vb A_\lambda (\vb k)=0$, and derive Eq.~\eqref{isz2} in this gauge in the following.

A bosonic BdG system is said to respect IS, if there exists an inversion operator $\mathcal P$, such that
\begin{equation}
	\mathcal P H^\eff_{\vb k} \mathcal P\inv = H^\eff_{-\vb k}.\nonumber
\end{equation}
Here, $\mathcal P$ is assumed to be independent of $\vb k$, \textit{pseudo-unitary}, \textit{pseudo-Hermitian} [cf. Eq.~\eqref{prfcuni} and \eqref{prfcansy}], square to $+1$, and commute with $\mathcal T$. By definition, we have
\begin{equation}
	\mathcal P\mathcal T H^\eff_{\vb k} (\mathcal P\mathcal T)\inv = H^\eff_{\vb k},\nonumber
\end{equation}
i.e., all energy bands are at least doubly degenerate at each $\vb k$. It is also straightforward to show that at PTRIM, each Kramers' pair have the same inversion eigenvalue, hence there is no ambiguity in choosing which one of the parity eigenvalue for the $\lambda $-th pair in Eq.~\eqref{isz2}. Another immediate fact is that the Berry curvature $F_\lambda (\vb k)$ must vanish since it is both odd and even in $\vb k$ due to PTRS and IS, respectively. Now define the unitary and antisymmetric sewing matrix $C$ in an arbitrary gauge by
\begin{equation}
	\ket{u_m(\vb k)}= -C^*_{mn}(\vb k)\mathcal P\mathcal T\ket{u_n(\vb k)}.\label{sewmatcdef}
\end{equation}
Assuming no other degeneracies, it is a $2\times 2$ matrix labeled by $\lambda $. The Pfaffian of $C_\lambda (\vb k)$ has unit magnitude and the gradient of its phase is related to the Berry connection Eq.~\eqref{berryconnection} by \cite{supp},
\begin{equation}\label{gfusingc}
	\begin{split}
		\vb A_\lambda (\vb k) &=-\frac{i}{2}\tr[C^\dagger_\lambda (\vb k) \nabla_{\vb k}C_\lambda (\vb k)]\\
		&=-i \nabla_{\vb k}\log \pf[C_\lambda (\vb k)].
	\end{split}
\end{equation}
By setting $\pf[C_\lambda (\vb k)]=1$ via a suitable gauge transformation, the Berry connection vanishes. Due to \cite{supp}
\begin{equation}\label{cbrelation}
	C_\lambda (-\vb k)=B_\lambda (\vb k)C^*_\lambda (\vb k) B^T_\lambda (\vb k),
\end{equation}
and $\pf[XAX^T]=\pf[A]\det[X]$, this gauge also guarantees that $\det[B_\lambda (\vb k)]=1$. Finally, from Eq.~\eqref{smb}, we have
\begin{equation}\label{bc1}
	[B(\vb \Lambda _i)]_{ll'}=\krin{\psi_\lambda ^{(l)}(\vb \Lambda _i),\mathcal P(\mathcal P\mathcal T)\psi_\lambda ^{(l')}(\vb \Lambda _i)}
\end{equation}
where $\ket{\psi_\lambda ^{(l)}(\vb \Lambda _i)}=e^{i\vb \Lambda _i\cdot\vb r}\ket{u_\lambda ^{(l)}(\vb \Lambda _i)}$ is the Bloch eigenstate. Since $[H,\mathcal P]=0$, $\ket{\psi_\lambda(\vb \Lambda _i)}$ is also the eigenstate of $\mathcal P$ with eigenvalue $\xi_\lambda =\pm 1$. Then using Eq.~\eqref{cdefanother}, Eq.~\eqref{bc1} leads to $B_\lambda (\vb \Lambda _i)=\xi_\lambda (\vb \Lambda _i)C_\lambda (\vb \Lambda _i)$, which gives $\pf[B_\lambda (\vb \Lambda _i)]=\xi_\lambda (\vb \Lambda _i)\pf[C_\lambda (\vb \Lambda _i)]=\xi_\lambda (\vb \Lambda _i)$ in the transverse gauge. All in all, we have $\sqrt{\det[B_\Lambda (\vb \Lambda _i)]}/\pf[B_\Lambda (\vb \Lambda _i)] =\xi_\Lambda (\vb \Lambda _i)$, and Eq.~\eqref{isz2} then follows from Eq.~\eqref{z2dl}.

Finally, we note that according to the bulk-boundary correspondence, the helical mid-gap edge states crossing at PTRIM is present (absent) if $\mathbb Z_2$ index equals one (zero). For the case that the gap is between $n$-th and $(n+1)$-th particle bands, the corresponding $\mathbb Z_2$ index is obtained by considering all bands below the $n+1$-th band, \textit{including all hole bands}. Since it has been shown that a pair of particle bands and its hole companion have the same $\mathbb Z_2$ index, one can equivalent sum all contributions \textit{above} the $n$-th particle bands. An additional merit of this treatment is that we hence avoid the ambiguity relating to the lowest particle bands (and highest hole bands), where the presence of Goldstone modes leads to undefined points in the Brillouin zone \cite{Furukawa2015}.


\begin{figure}
	\includegraphics[width=8.0cm]{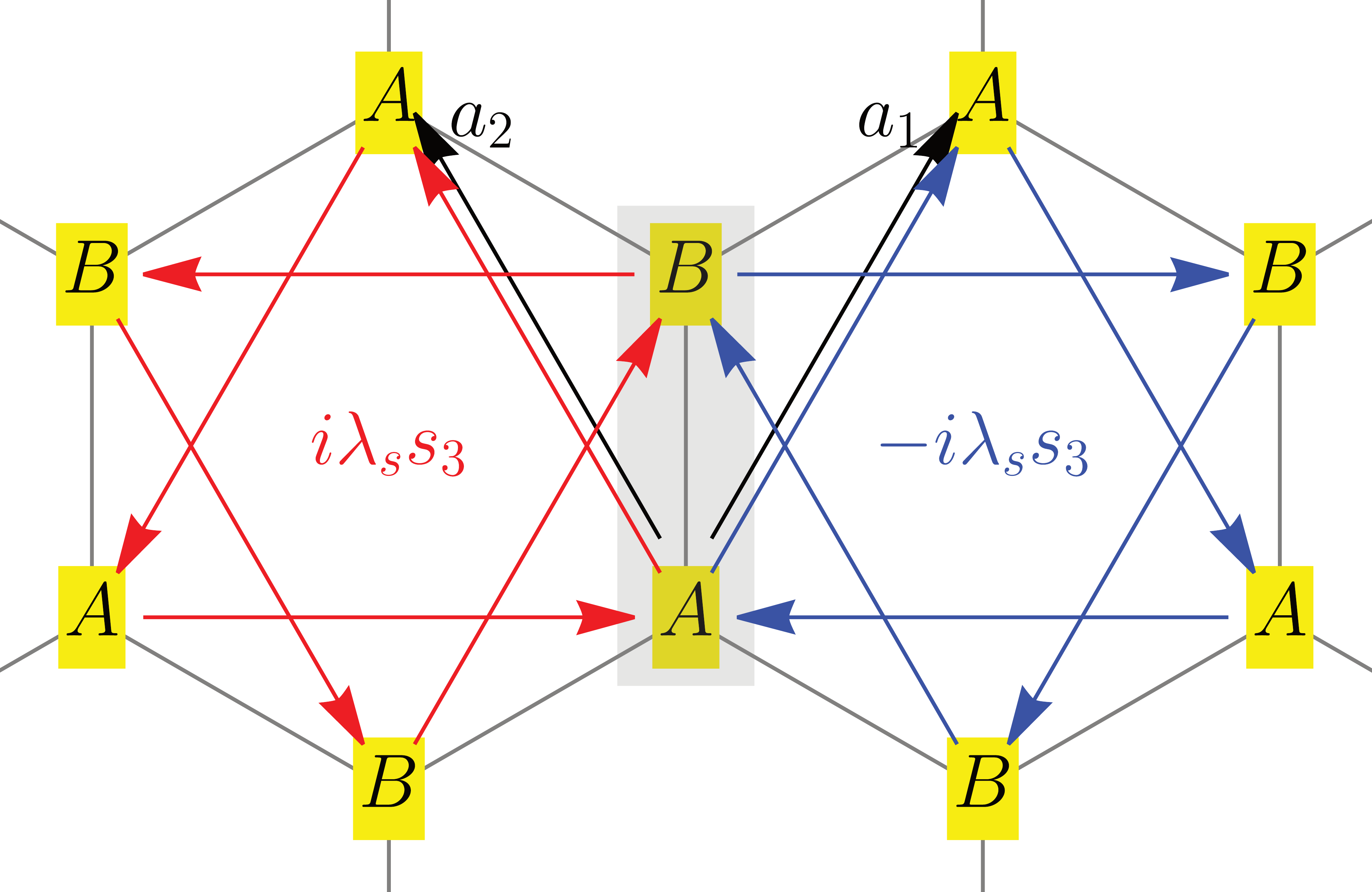}
	\caption{\label{bkmlattice}Kane-Mele model on a honeycomb lattice. Two sublattices, highlighted in yellow, are denoted as $A$ and $B$, with a unit cell indicated by the gray rectangle. $\vb a_1$ and $\vb a_2$ are the Bravais lattice vectors. The gray line denotes the NN hopping amplitude $-t$, while the NNN hopping matrix of Haldane's type are shown in red and blue with the arrow implying the hopping direction. Note the relative minus sign between two species of bosons are captured by the standard Pauli matrix $s_3$ acting on the pseudospin space.}
\end{figure}


\section{Toy models}\label{sec3}
We examine the topological properties of excitation spectrum by calculating the bulk $\mathbb Z_2$ invariant in two ways (as an obstruction and as the Wannier center flow), and numerically verify the bulk-boundary correspondence for two toy models that are feasible in cold atom systems: the bosonic version of (1) Kane-Mele model and (2) Bernevig-Hughes-Zhang model. The former has PTRS but breaks IS in general, while the latter always preserves both. When IS exists, We also explicitly verify Eq.~\eqref{isz2}.

\begin{figure*}
	\includegraphics[width=17cm]{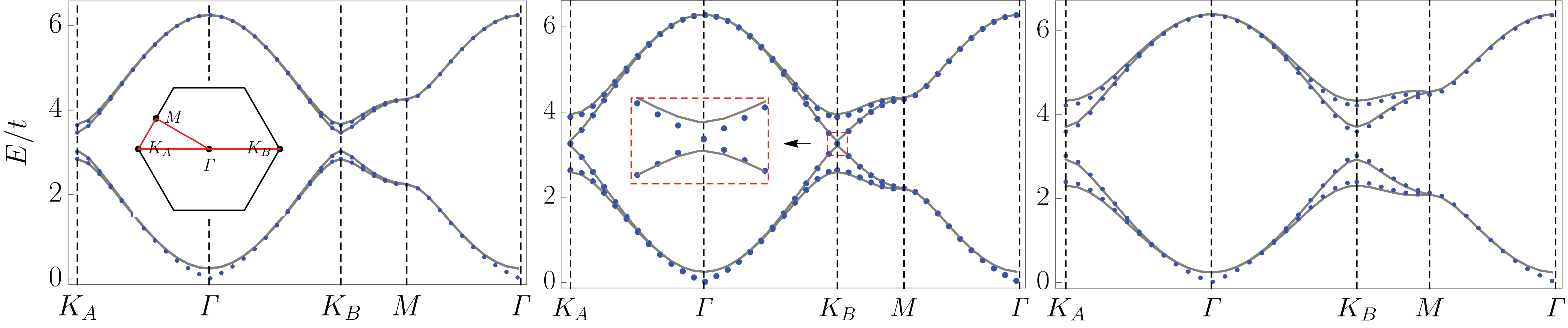}
	\caption{\label{bkmbe}Bogoliubov excitation spectrum (blue dots) under full periodic boundary conditions along high symmetric lines in the 1BZ (shown in the inset of the left figure), for (left) $\lambda_v/t=0.10$, (middle) $\lambda_v/t=\lambda_v^\star/t \approx 0.36$ and (right) $\lambda_v/t=0.70$. The noninteracting bands (shifted upwards by $nU/2-\mu$) are also plotted in gray solid lines. As shown in the inset of the middle figure, when the gap of Bogoliubov excitation spectrum just closes, the corresponding gap of the noninteracting band has already reopened again. Other relevant parameters: $nU/t=1$ and $\lambda _s/t=0.06$.}
\end{figure*}

\subsection{Bosonic Kane-Mele model}
Our first example is a bosonic version of the Kane-Mele (BKM) model \cite{Kane2005}, which is a time-reversal-symmetric generalization of Haldane's honeycomb lattice model \cite{Haldane1988}. Since the latter has been realized in a cold atom experiment by Esslinger's group \cite{Jotzu2014}, we expect this model is ready for implementation. The noninteracting part of the Hamiltonian reads
\begin{equation}\label{bkmh0}
	H_0 = -t\sum_{\expval{i,j}}a^\dagger_i a_j-i\lambda _s\sum_{\krin{i,j}} v_{ij} a^\dagger_i s_3 a_j - \lambda_v \sum_i \xi_i a^\dagger_i a_i,
\end{equation}
where $a^{(\dagger)}_i$ is the bosonic annihilation (creation) operators at site $i$, with pseudospin index omitted. The first term describes the nearest-neighbor hopping; the second term describes the next-nearest-neighbor (NNN) complex hopping for both pseudospin sectors with an overall opposite sign between them (see Fig.~\ref{bkmlattice}). Here $s_i$ ($s_0$), $i=1,2,3$, is the standard Pauli (two-by-two identity) matrix acting on the pseudospin space. $v_{ij}=\sgn(\vb d_i\times \vb d_j)_z=\pm 1$ with $\vb d_i$ and $\vb d_j$ along the two bonds constituting the next-nearest neighbors. The last term is a staggered sublattice potential $\xi_i=1 (-1)$ for $i\in  A (B)$, which breaks IS of the system. The interacting part takes the form
\begin{equation}
	H_{\tint}= \frac{U}{2}\sum_{j}\sum_{s=\uparrow,\downarrow} a^\dagger_{js}a^\dagger_{js}a_{js} a_{js},\label{hint}
\end{equation}
where the interspecies interactions are neglected for simplicity.

Using primitive lattice vector as shown in Fig.~\ref{bkmlattice}, we write Eq.~\eqref{bkmh0} in momentum space as $H_0=\sum_{\vb k}a^\dagger_{\vb k}h(\vb k) a_{\vb k}$, where $a_i \coloneqq (a_{iA\uparrow},a_{iA\downarrow},a_{iB\uparrow},a_{iB\downarrow})^T$ and the 4-by-4 Bloch Hamiltonian reads,
\begin{equation}\label{bkmh0k}
	h(\vb k)=d_1(\vb k)\Gamma_1+d_2 \Gamma_2+d_{12}(\vb k)\Gamma_{12}+d_{15}(\vb k)\Gamma_{15},
\end{equation}
with the Clifford algebra generators
\begin{equation}
	\Gamma_a=(\sigma_1 \otimes s_0,\sigma_3 \otimes s_0,\sigma_2\otimes s_1, \sigma_2\otimes s_2,\sigma_2\otimes s_3),\label{gammabkm}
\end{equation}
and $\Gamma_{ab}=\frac{1}{2i}[\Gamma_a,\Gamma_b]$, where $\sigma_i$ ($\sigma_0$), $i=1,2,3$, are Pauli (two-by-two identity) matrix acting on the sublattice space. All real parameters are listed in Table~\ref{paramtab}. In this representation we have $\tilde{\mathcal T} \Gamma_a \tilde{\mathcal T}\inv =\Gamma_a$ and $\tilde{\mathcal T} \Gamma_{ab} \tilde{\mathcal T}\inv =-\Gamma_{ab}$, with
\begin{equation}
	\tilde{\mathcal T}=i(\sigma_0\otimes s_2) K = -i\Gamma_{35}K.\label{PTRSop1}
\end{equation}
Hence $d_{a}$ ($d_{ab}$) is even (odd) in $\vb k$ dictated by the odd TRS: $\tilde{\mathcal T}h(\vb k)\tilde{\mathcal T}\inv = h(-\vb k)$.

\begin{table}
	\caption{\label{paramtab}Parameters used in Eq.~\eqref{bkmh0k} are given in the first two rows, and extra parameters used in Eq.~\eqref{bkmheff} are given in the last two rows, with $x=\sqrt{3} k_x a/2$ and $y=3 k_y a/2$. Note $\bar \theta$ itself is a function of $\lambda_v/t$ and $Un/t$, see Fig.~\ref{bkmtheta}.}
\begin{ruledtabular}
	\begin{tabular}{cccc}
		$d_1$ & $-t(1+2\cos x\cos y)$ & $d_2$ & $-\lambda_v$ \\
		$d_{12}$ & $2t\cos x\sin y$ & $d_{15}$ & $4\lambda_s(\cos x-\cos y)\sin x$\\
		\hline
		$d_0$ & $\lambda_v\cos \bar \theta+3t\sin \bar \theta+\frac{1}{4}nU\sin^2 \bar \theta$ & $\tilde d_0$ & $\frac{1}{4}nU$ \\
		$\tilde d_2$ & $-\lambda_v+\frac{1}{2}nU \cos \bar \theta$ & $\tilde{\tilde{d}}_2$ & $\frac{1}{4}nU \cos \bar \theta$
	\end{tabular}
\end{ruledtabular}
\end{table}

\begin{figure}
	\includegraphics[width=8.4cm]{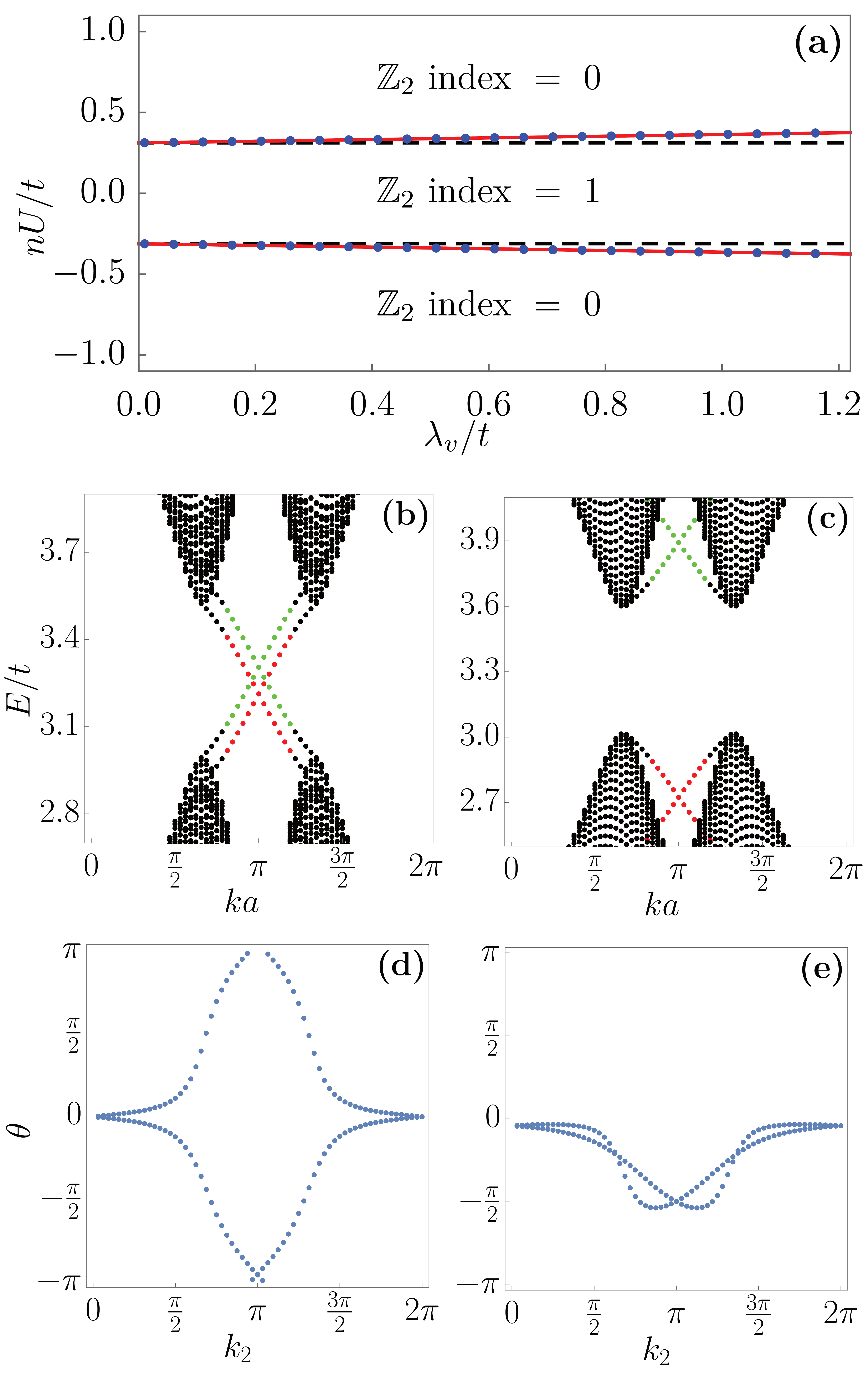}
	\caption{\label{bkmtpd}\textbf{(a)} $\mathbb Z_2$ index, calculate using Eq.~\eqref{fukui} and the numerical method of \cite{Fukui2007}, for the higher two particle bands in the BKM model with $nU/t=1$ and $\lambda_s/t=0.06$. The boundaries, shown in blue dotted lines, are obtained via a full numerical calculations [explained above Eq.~\eqref{lv}]. Red solid lines are obtained from the series expansion, Eq.~\eqref{lv}. Comparing with the black dashed vertical lines (corresponding to the noninteracting case) shows that the $\mathbb Z_2$ topological region becomes larger. In \textbf{(b)} $\lambda_v/t=0.1$ and \textbf{(c)} $\lambda_z/t=0.7$, we zoom in around the mid-gap of the Bogoliubov excitation spectrum of particle bands in a strip geometry of $64$ unit cells (each with $64$ sites) with zigzag edges. Red (blue) points corresponding to edge modes, whose wavefunctions have more than $80\%$ weight on the leftmost (rightmost) unit cell. \textbf{(d)} and \textbf{(e)} are the corresponding Wannier center flow of the second pair of particle bands by treating $k_2$ as the time.}
\end{figure}

The single-particle band minimum obtained from Eq.~\eqref{bkmh0k} is the same as the Haldane model studied by Ref.~\cite{Vasic2015,Furukawa2015}: For $\abs{\lambda _s}<t/\sqrt{3}$, the band bottom locates at $\vb{\Gamma}$; while for larger $\abs{\lambda _s}$, the minimum jumps to the corners $\vb K_A$ and $\vb K_B$ of the first Brillouin zone (1BZ). Here we focus on the former case where the condensation is expected to occur at $\vb \Gamma$. Then, by taking a general ground state wave function ansatz and minimizing the corresponding Gross-Pitaevskii (GP) energy functional \cite{supp}, the superfluid order parameter is found to be $\expval{a_{iA\uparrow}}=\expval{a_{iA\downarrow}}=\cos(\bar \theta/2)\sqrt{n}/2$ and $\expval{a_{iB\uparrow}}=\expval{a_{iB\downarrow}}=\sin(\bar \theta/2)\sqrt{n}/2$ with $\bar \theta$ as function of $\lambda_v/t$ and $Un/t$ [cf. Fig.~\ref{bkmtheta}].

Applying the standard Bogoliubov theory \cite{supp}, we obtain the effective Hamiltonian (note it is non-Hermitian due to the presence of imaginary $i$ in the first line on the r.h.s.),
\begin{equation}\label{bkmheff}
	\begin{split}
		H^\eff_{\vb k} &= \tau_0 \otimes [d_{15}(\vb k) \Gamma_{15}] +i \tau_2 \otimes (\tilde d_0 \Gamma_0 + \tilde{\tilde{d}}_2 \Gamma_2)\\
		&\quad + \tau_3 \otimes [d_0\Gamma_0 + d_1(\vb k) \Gamma_1 + \tilde{d}_2 \Gamma_2 + d_{12}(\vb k) \Gamma_{12}]
	\end{split}
\end{equation}
where $\Gamma_0=\sigma_0\otimes s_0$ and all real parameters are listed in Table~\ref{paramtab}. One can easily see that this effective Hamiltonian has PTRS with the PTR operator
\begin{equation}
	\mathcal T= \tau_0\otimes\tilde{\mathcal T}.\nonumber
\end{equation}
Diagonalization of Eq.~\eqref{bkmheff} leads to the Bogoliubov excitation spectrum shown in Fig.~\ref{bkmbe} along the high symmetric lines. In the low energy limit, there are two Goldstone modes resulting from the spontaneous breaking of two $U(1)$ symmetries corresponding to particle number and $s_3$ conservations, respectively.

For the noninteracting Hamiltonian, Eq.~\eqref{bkmh0k}, the system is gapped for a general $\lambda_v$ at half filling. The gap-closing-and-reopening transition occurs at two corners $\vb K_A$ and $\vb K_B$ when $\lambda_v=\pm 3\sqrt{3}\lambda_s$ \cite{Kane2005}. This behavior is smoothly carried over to the Bogoliubov excitation spectrum, with the only difference that now $\lambda_v=\lambda_v^\star\neq \pm 3\sqrt{3}\lambda_s$. By calculating eigenvalues of $H_{\vb K_A}^\eff$ (or equivalently $H_{\vb K_B}^\eff$) and equating the second and third eigenvalues, the critical value $\lambda_v^\star$ is found to be a function of $\lambda_v/t$ and $nU/t$, which is plotted as blue dots in Fig.~\ref{bkmtpd}(a). For the case with $\lambda_s,nU\ll t$, it takes the form
\begin{equation}\label{lv}
	\lambda_v^\star \approx 3\sqrt{3}\lambda _s+\frac{\sqrt{3}}{2}\lambda _s nU/t,
\end{equation}
which indeed returns to the noninteracting case by setting $U=0$. Using Eq.~\eqref{fukui} with the numerical method of Fukui and Hatsugai \cite{Fukui2007}, for the higher pair of particle bands, we find $\tilde \Delta_2=1$ for $\abs{\lambda_v}<\lambda_v^\star$, corresponding to the $\mathbb Z_2$ topological region; and $\tilde \Delta_2=0$ for $\abs{\lambda_v}>\lambda_v^\star$, corresponding to the $\mathbb Z_2$ trivial region. We also calculate the Wannier center flow for the higher pair of particle bands [Fig.~\ref{bkmtpd}(d,e)], and the Bogoliubov excitations in a one-dimensional zigzag strip [Fig.~\ref{bkmtpd}(b,c)], which confirms the equivalence of two definitions of the $\mathbb Z_2$ invariant, and the bulk-boundary correspondence (i.e., the presence/absence of helical edge states for $\mathbb Z_2$ topological/trivial region).

For $\lambda_v=0$, we have $\bar \theta = \pi/2$, and the effective Hamiltonian enjoys IS with the inversion operator $\mathcal P=\tau_0\otimes\Gamma_1$. At four PTRIM, we find,
\begin{equation*}
	\begin{split}
		(\xi_{00},\xi_{01},\xi_{10},\xi_{11})=(-1,-1,-1,1),
	\end{split}
\end{equation*}
where $\xi_{ij}$ is the eigenvalue of $\mathcal P$ for the second pair of particle bands at PTRIM $\vb k = i \vb b_1/2 +j \vb b_2/2$. Eq.~\eqref{isz2} then indicates that the case with $\lambda_v=0$ is in the $\mathbb Z_2$ topological region, as expected.

As seen both from Eq.~\eqref{lv} (in a specific limit) and from Fig.~\ref{bkmtpd}(a), the $\mathbb Z_2$ topological region becomes larger with increasing the interaction strength (or the particle number density). To understand why this is the case physically, we first note that the large $\lambda_v$ limit corresponds to the ``atomic limit" \cite{Bernevig2013}, where all atoms are tightly located at one of the two sublattices, corresponding to an extreme sublattice imbalance, and is of course $\mathbb Z_2$ trivial. By turning on a large $\abs{\lambda_v}$ from zero (i.e., from $\mathbb Z_2$ topological region), one encounters a gap-closing-and-reopening transition. Effects of the repulsive interaction, on the other hand, suppress the sublattice imbalance induced by $\lambda_v$, since it favors a uniform configuration. As a result, to reach the critical value of sublattice imbalance, one needs a larger $\lambda_v$.
\subsection{Bosonic Bernevig-Hughes-Zhang model}
\begin{figure}
	\includegraphics[width=8.4cm]{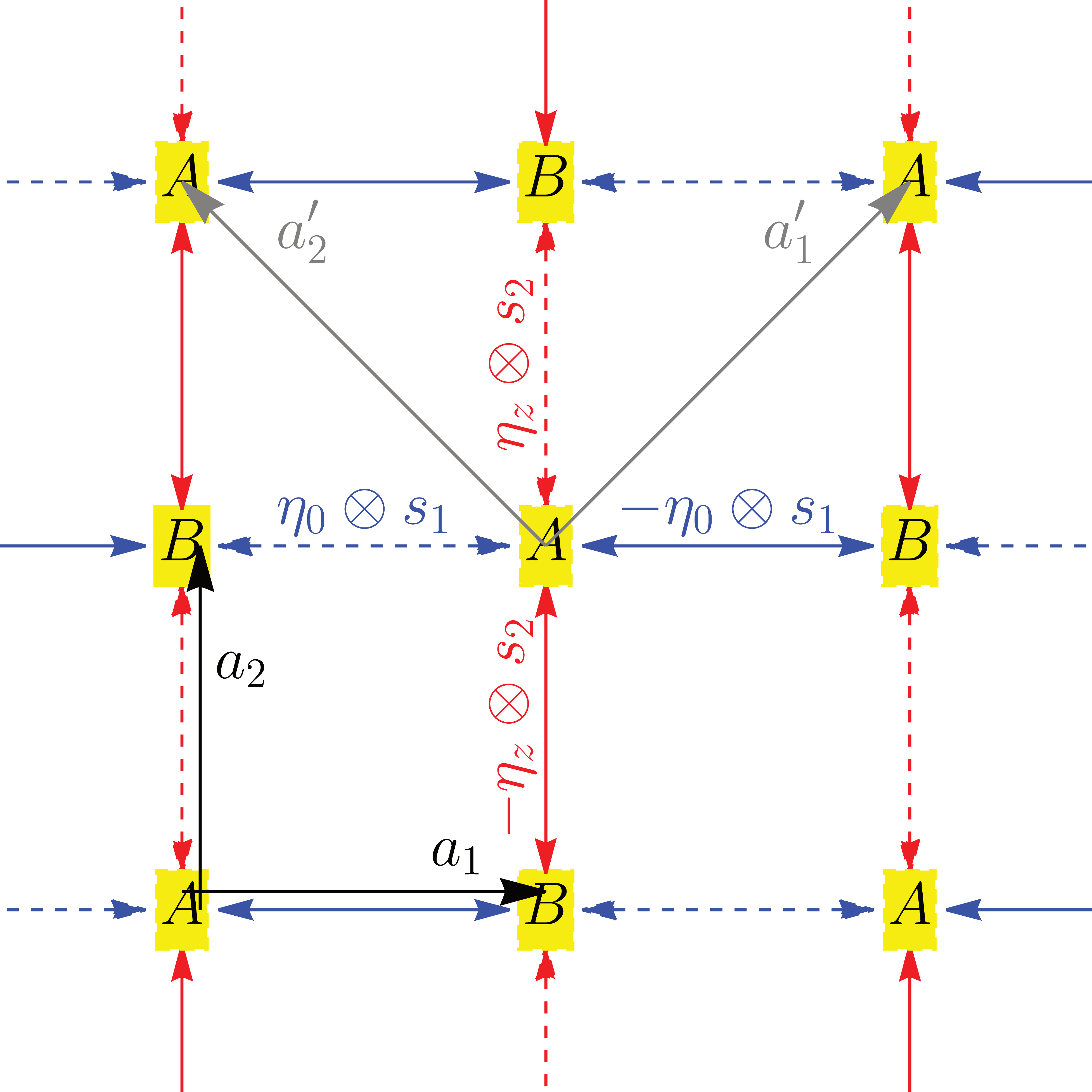}
	\caption{\label{bbhzlattice}Bernevig-Hughes-Zhang-like model on a square lattice. It is a straightforward generalization of the scheme proposed by Liu \etal \cite{Liu2014}. The pseudospin-flip hopping matrix $h_s^{ij}$ along x (y) direction, used in Eq.~\eqref{bhzorigin}, are shown in blue (red), with the arrow indicating the hopping direction. There is a relative minus sign between dashed lines and solid lines of the same color. Hence in the original gauge, the unit cell contains two sites (denoted as $A$ and $B$ with yellow background), with the corresponding Bravais primitive lattice vector $\vb a_{1,2}'$ shown in gray \cite{Lang2017}. Only \textit{after} performing a clever gauge transformation $a_{j\eta\downarrow}\rightarrow (-1)^{j}a_{j\eta\downarrow}$ \cite{Liu2014}, all lines become solid, and it is then valid to use $\vb a_{1,2}$ shown in black, as the two Bravais lattice vectors, and the unit cell contains only a single site.}
\end{figure}

%
%
Our second example is a bosonic version of the Bernevig-Hughes-Zhang (BBHZ) model, which is a time-reversal-symmetric generalization of the Chern insulator on the square lattice. Motivated by a scheme proposed by Liu \etal \cite{Liu2014} which has been experimentally realized by Pan's group \cite{Wu_2016,Sun2018}, we consider two copies (labeled by $\eta=A,B$) of pseudospin-1/2 (labeled by $s=\uparrow,\downarrow$) bosons on the square lattice, with the tight-binding Hamiltonian (Fig.~\ref{bbhzlattice}),
\begin{equation}\label{bhzorigin}
\begin{split}
	H_0 &=-t\sum_{\expval{i,j}}a^\dagger_i (\eta_0\otimes s_0) a_j-m_z\sum_i a^\dagger_i (\eta_0\otimes s_3) a_i  \\
	&\quad -t_s \sum_{\expval{i,j}} a^\dagger_i h_s^{ij} a_j,
\end{split}
\end{equation}
where $a_{i\eta s}^{(\dagger)}$ is the annihilation (creation) operator of $\eta$ boson with pseudospin $s$ at site $\vb r_i$, and $a_i \coloneqq (a_{iA\uparrow},a_{iA\downarrow},a_{iB\uparrow},a_{iB\downarrow})^T$. $\eta_i$ ($\eta_0$) and $s_i$ ($s_0$), $i=1,2,3$, are Pauli (two-by-two identity) matrix acting on boson-copy space and pseudospin space, respectively.  $t$ denotes pseudospin-conserved hopping, $m_z$ is a constant Zeeman term. The pseudospin-flip hopping matrix $h_s^{ij}$ are shown explicitly in Fig.~\ref{bbhzlattice}. Note the relative minus sign, i.e., the presence of $\eta_3$, between two copies of bosons when they hop along the $x$ direction, makes Eq.~\eqref{bhzorigin} odd time-reversal symmetric  \cite{Asboth2016}. The interacting part of the Hamiltonian takes the same form as Eq.~\eqref{hint} for each copy of boson.

After a gauge transformation $a_{j\eta\downarrow}\rightarrow (-1)^{j}a_{j\eta\downarrow}$ \cite{Liu2014}, the momentum-space Bloch Hamiltonian is
\begin{equation}\label{hkbhz}
\begin{split}
		h(\vb k)&=\eta_0\otimes\big\{[-2t(\cos k_x+\cos k_y)-m_z] s_3\\
	&\quad -2t_s \sin k_x s_2\}+\eta_3\otimes(-2t_s \sin k_y s_1),
\end{split}
\end{equation}
which resembles the 4-band model for HgTe introduced by Bernevig, Hughes and Zhang \cite{Bernevig2006}. This single-particle Hamiltonian has both odd TRS
\begin{equation}
	\tilde{\mathcal T}h(\vb k)\tilde{\mathcal T}\inv =h(-\vb k),\quad \tilde{\mathcal T}=i\eta_2\otimes s_0 K,\nonumber
\end{equation}
and IS
\begin{equation}
	\tilde{\mathcal P}h(\vb k)\tilde{\mathcal P}\inv =h(-\vb k),\quad \tilde{\mathcal P}=\eta_0\otimes s_3.\nonumber
\end{equation}
Instead of Eq.~\eqref{gammabkm}, it is convenient to choose the Dirac matrices to be even under $\tilde{\mathcal P}\tilde{\mathcal T}$ \cite{Fu2007},
\begin{equation*}
	\Gamma_a=\pqty{\eta_0\otimes s_3,\eta_0\otimes s_2,\eta_1\otimes s_1,\eta_2\otimes s_1,\eta_3\otimes s_1}.
\end{equation*}
Then $\tilde{\mathcal T}=-i\Gamma_{35}K$, $\tilde{\mathcal P}= \Gamma_1$, and the commutators are odd under $\tilde{\mathcal P}\tilde{\mathcal T}$, i.e., $\tilde{\mathcal P}\tilde{\mathcal T}\Gamma_{ab}(\tilde{\mathcal P}\tilde{\mathcal T})\inv=-\Gamma_{ab}$. Due to $\Gamma_1=\tilde{\mathcal P}$, we further have
\begin{equation}
	\tilde{\mathcal T} \Gamma_a\tilde{\mathcal T}= \tilde{\mathcal P}\Gamma_a\tilde{\mathcal P}=\begin{cases}
		&+\Gamma_a \qfor a=1,\\
		&-\Gamma_a \qfor a\neq 1.
	\end{cases}\nonumber
\end{equation}
Eq.~\eqref{hkbhz} is then recast into
\begin{equation}\label{hkbhz1}
	h(\vb k)=d_1(\vb k)\Gamma_1+d_{2}(\vb k)\Gamma_{2}+d_{5}(\vb k)\Gamma_{5},
\end{equation}
with all real coefficients listed in Table~\ref{paramtabBHZ}. As shown in Ref.~\cite{Fu2007}, the $\mathbb Z_2$ invariant for this noninteracting model can be identified using the same formula as Eq.~\eqref{isz2}. The representation of Dirac matrices has been chosen such that at the four TRIM, only $d_1$ can be nonzero, i.e., $h(\vb \Lambda )=d_1(\vb \Lambda )\tilde{\mathcal P}$. Hence one can directly obtain all eigenvalues of $\tilde{\mathcal P}$ for the occupied bands,
\begin{equation}
	(\xi_{00},\xi_{01},\xi_{10},\xi_{11})=(-4t-m_z,-m_z,-m_z,4t-m_z),\label{nonBHZis}
\end{equation}
which shows that the noninteracting system is in the $\mathbb Z_2$ topological region for $\abs{m_z}<4t$, and in the $\mathbb Z_2$ trivial region otherwise.

\begin{table}
	\caption{\label{paramtabBHZ}Parameters used in Eq.~\eqref{hkbhz1} are given in the first two rows, and extra parameters used in Eq.~\eqref{bbhzheff} are given in the last two rows.}
\begin{ruledtabular}
	\begin{tabular}{cccc}
		$d_1$ & $-2t(\cos k_x +\cos k_y)-m_z$ & $d_{2}$ & $-2t_s\sin k_x$ \\
		$d_{5}$ & $-2t_s \sin k_y$ & \\
		\hline
		$\tilde d_1$ & $\frac{nU}{2}+d_1$ & $d_0$ & $\frac{nU}{4}$ \\
		$\tilde d_0$ & $4t+m_z$ &  & 
	\end{tabular}
\end{ruledtabular}
\end{table}

Single-particle bands of Eq.~\eqref{hkbhz} are the same as the Chern insulator with a two-fold degeneracy dictated by $\tilde{\mathcal P}\tilde{\mathcal T}$ symmetry. For $\abs{t_s}<t_s^\star$, where $t_s^\star=\sqrt{2t^2+m_z t/2}$, the band minimum locates at $\vb \Gamma$ ($\vb M$) if $m_z>0$ ($m_z<0$). For $\abs{t_s}>t_s^\star$, this single minimum splits into four points, $(\pm k_0,\pm k_0)$ with $k_0=\arccos\frac{m_z t}{2(t_s^2 -2t^2)}$. Here we will focus on the former case, and assume bosons condense \textit{only} at $\vb \Gamma$, by taking a sufficiently large positive $m_z$. Then, using a general ground state wave function ansatz and minimizing the corresponding GP energy functional \cite{supp}, the superfluid order parameter is found to be $\expval{a_{iA\uparrow}}=\expval{a_{iB\uparrow}}=1/\sqrt{2}$ and $\expval{a_{iA\downarrow}}=\expval{a_{iB\downarrow}}=0$, for the physically relevant region (cf. Fig.~\ref{bbhztheta}).

\begin{figure*}
	\includegraphics[width=17cm]{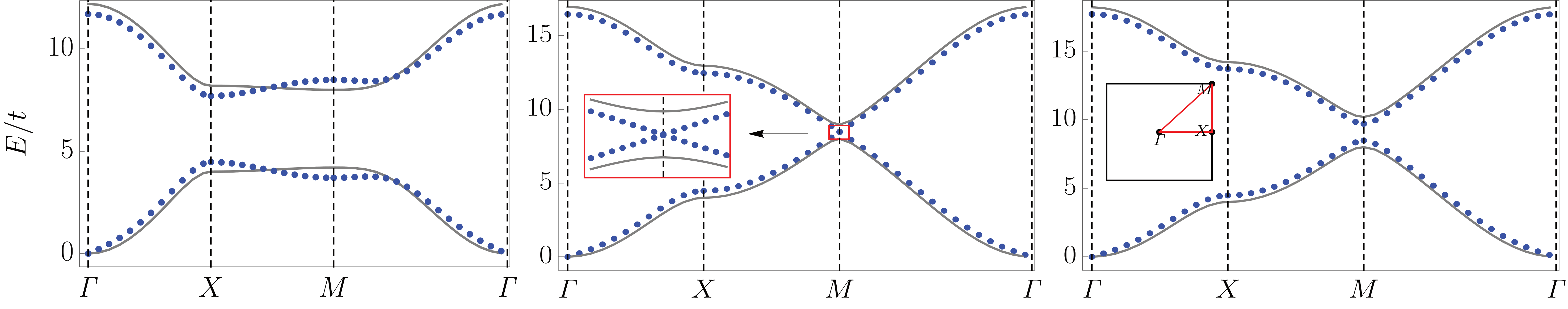}
	\caption{\label{bbhzbe} Bogoliubov excitation spectrum (blue dots) under full periodic boundary conditions along high symmetric lines in the 1BZ (shown in the inset of the right figure), for (left) $m_z/t=2.10$, (middle) $m_z/t=m_z^\star/t\approx 4.48$ and (right) $m_z/t=5.10$. The noninteracting bands (shifted upwards by $nU/2-\mu$) are also plotted in gray lines. As shown in the inset of the middle figure, when the gap of Bogoliubov excitation spectrum just closes, the gap corresponding to the noninteracting band has already reopened. Other relevant parameters: $nU/t=t_s/t=1$.}
\end{figure*}

Using the Bogoliubov theory, we obtain the effective Hamiltonian (note it is non-Hermitian due to the presence of imaginary $i$ in the first line on the r.h.s.) \cite{supp},
\begin{equation}\label{bbhzheff}
	\begin{split}
		H_{\vb k}^\eff &= d_{5}\tau_0\otimes\Gamma_{5} +id_0 \tau_2\otimes(\Gamma_0+\Gamma_1)\\
		&\quad \tau_3\otimes(\tilde{d}_0 \Gamma_0 + \tilde{d}_1 \Gamma_1 + d_2 \Gamma_{2}),
	\end{split}
\end{equation}
with $\Gamma_0=\eta_0\otimes s_0$ and all real parameters listed in Table~\ref{paramtabBHZ}. It is straightforward to check that this effective Hamiltonian has both PTRS with the PTR operator
\begin{equation}
	\mathcal T=\tau_0\otimes\tilde{\mathcal T},\label{PTRSop2}
\end{equation}
and IS with the inversion operator
\begin{equation}
	\mathcal P=\tau_0\otimes\tilde{\mathcal P}.\label{ISop}
\end{equation}
Diagonalization of Eq.~\eqref{bbhzheff} leads to the Bogoliubov excitation spectrum shown in Fig.~\ref{bbhzbe}, which is doubly degenerate dictated by $\mathcal P\mathcal T$ symmetry. In the low energy limit, there are two Goldstone modes due to the spontaneous breaking of two $\mathrm{U}(1)$ symmetries associated to particle number and $\eta_3$ conservations, respectively.

\begin{figure}
	\includegraphics[width=8.4cm]{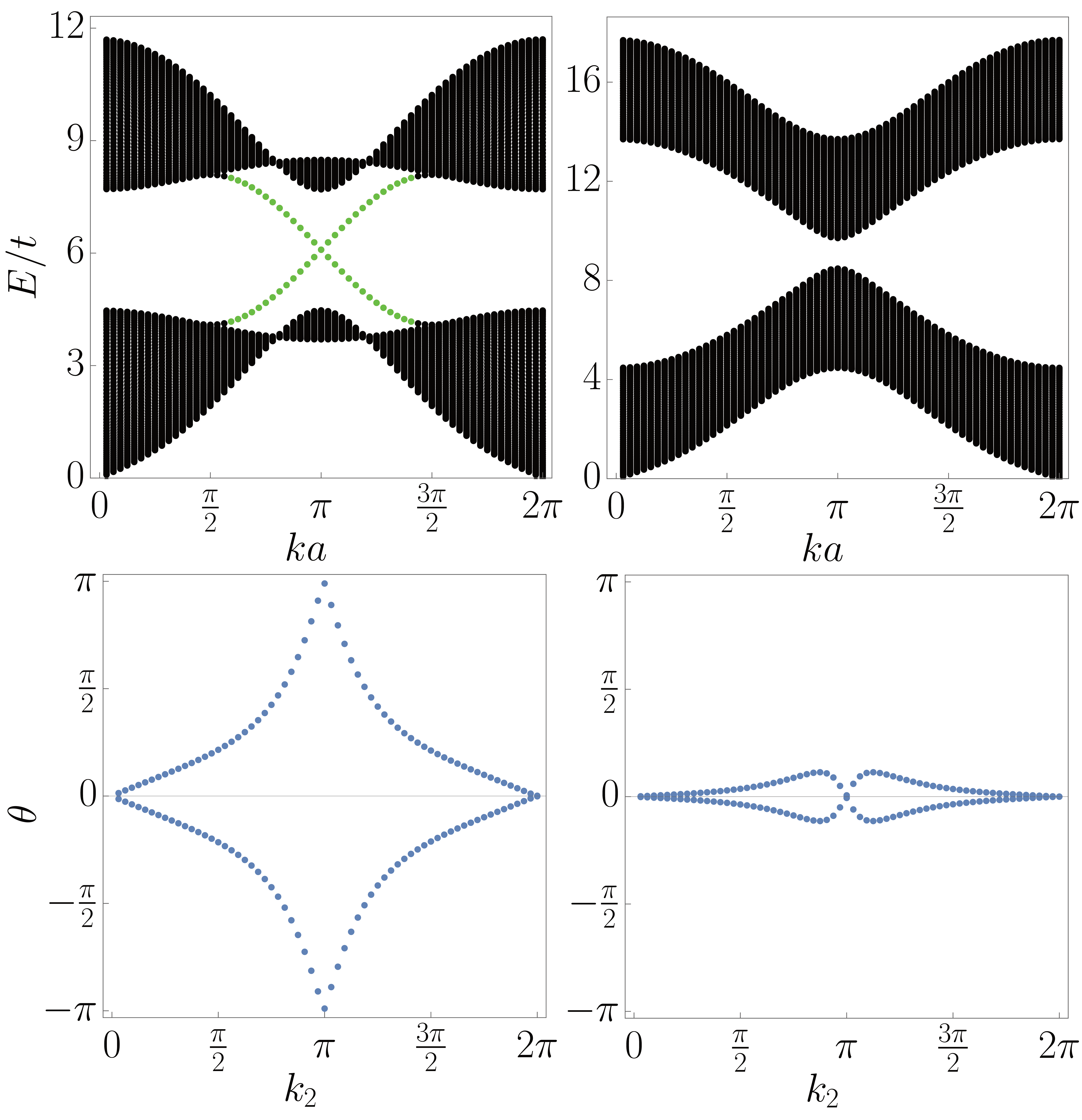}
	\caption{\label{bbhztpd} Bogoliubov excitation spectrum of particle bands in a strip geometry of $64$ unit cells (each containing $64$ sites) for $m_z/t=2.1$ (left top) and $m_z/t=5.1$ (right top). Green points correspond to edge modes, whose wavefunctions have more than $80\%$ weight on the rightmost unit cell. Edge modes on the leftmost unit cell completely overlaps with the green points due to IS. The corresponding Wannier center flow of the second pair of particle bands by treating $k_2$ as the time are shown in bottom. Other relevant parameters: $nU/t=t_s/t=1$.}
\end{figure}

For the noninteracting Hamiltonian Eq.~\eqref{hkbhz1}, the gap-closing-and-reopening transition occurs at $\vb M$ when $m_z=4t$ for $m_z$ positive [cf. Eq.~\eqref{nonBHZis}]. This behavior is again smoothly carried over to the Bogoliubov excitation spectrum, with the only difference that now $m_z=m_z^\star\neq4t$. By calculating eigenvalues of $H_{\vb M}^\eff$ and equating the two relevant ones, we find
\begin{equation}\label{lzc}
	m_z^\star= \sqrt{2t(8t+nU)}+\frac{nU}{4},
\end{equation}
which indeed returns to the noninteracting case by setting $U=0$. Using Eq.~\eqref{fukui} with the numerical method of Fukui and Hatsugai \cite{Fukui2007}, for the higher pair of particle bands, we find $\tilde \Delta_2=1$ for $m_z<m_z^\star$, corresponding to the $\mathbb Z_2$ topological region, while $\tilde \Delta_2=0$ for $m_z>m_z^\star$, corresponding to the $\mathbb Z_2$ trivial region. We also calculate the Wannier center flow for the higher pair of particle bands, and the Bogoliubov excitations in a strip geometry as shown in Fig.~\ref{bbhztpd}, which confirms the equivalence of two definitions of the $\mathbb Z_2$ invariant, and the bulk-boundary correspondence.

Since this model always has IS, we may simply examine the parity eigenvalues at the PTRIM. Because the gap closes only at $\vb M$, the parity eigenvalues can only change there. We thus consider the eigenvalues of two relevant particle bands with the corresponding eigenstates at $\vb M$: (note $E_{l+1}=E_l$, for $l=1,2$)
\begin{equation*}
	\begin{split}
		E_1(\vb M) &=\sqrt{8t(8t+nU)},\qquad E_3(\vb M)=2m_z-\frac{nU}{2}, \\
		\ket{u_1(\vb M)} &\propto (0,0,-\frac{16t + nU +4\sqrt{2t(8t+nU)}}{nU},0,0,0,1,0)^T, \\ \ket{u_3(\vb M)} &= (0,0,0,1,0,0,0,0)^T.
	\end{split}
\end{equation*}
which are obviously also the eigenvectors of inversion operator with parity $1$ and $-1$, respectively. Hence the topological transition occurs at the degeneracy point $E_1(\vb M)=E_3(\vb M)$ which leads again to Eq.~\eqref{lzc}. To determine the presence or absence of helical edge modes between the first and second pair of particle bands, one has to find out all four eigenvalues of $\mathcal P$ for the second pair of bands at PTRIM,
\begin{equation*}
	(\xi_{00},\xi_{01},\xi_{10},\xi_{11})=(-1,1,1,1),\qfor m_z<m_z^\star.
\end{equation*}
Eq.~\eqref{isz2} then indicates that $m_z<m_z^\star$ corresponds to the $\mathbb Z_2$ topological region, while $m_z>m_z^\star$ is the $\mathbb Z_2$ trivial region, as expected.

Lastly, we note that the $\mathbb Z_2$ topological region of BBHZ model also becomes larger with increasing the interaction strength (or the particle number density), as seen from Eq.~\eqref{lzc}. Similarly to the BKM model, physically speaking, the repulsive interaction favors a uniform configuration, which suppresses the pseudospin imbalance induced by $m_z$. In turn, to reach the critical value of pseudospin imbalance, one needs a larger $m_z$, i.e., the $\mathbb Z_2$ topological region becomes larger.

\section{Conclusion and discussion}\label{sec5}
In this article, we studied topological Bogoliubov excitations in BEC in optical lattices protected by a PTRS that is analogous to topological insulators in class AII of fermions. The bulk topological $\mathbb Z_2$ invariant is shown to be characterized by the Pfaffian, the pseudo-time-reversal polarization, and the Wannier center flow. The last definition is most useful because it is gauge independent. With an additional inversion symmetry, this $\mathbb Z_2$ invariant can be identified by examining the inversion eigenvalues of the ``occupied" states at PTRIM. In two simple and experimentally feasible examples, we confirmed the bulk-boundary correspondence numerically, and found in both cases that the topological region is enlarged by the interaction or the particle number density, since the repulsive interaction favors a uniform configuration which suppresses the effects of sublattice (pseudospin) imbalance induced by $\lambda_v$ ($m_z$), which will lead a transition to the topological trivial region. Effectively, this topology becomes more ``robust".

Similarly to the fermionic case discussed in Ref.~\cite{ShunQingShen}, we expect that the topological properties discussed in this paper should be robust against weak disorder that (1) respect the PTRS and (2) sufficiently weak so that topological excitation band gap does not close and the system does not enter into the Bose glass phase (where the topology of excitation spectrum could change dramatically). Of course, a more serious study should be carried out in the future.

The bulk-boundary correspondence guarantees that a non-trivial $\mathbb Z_2$ index implies the existence of topological edge states which can be experimentally detected in cold atom experiments \cite{Goldman2012,Goldman2013,Goldman2013a,Celi2015}. We expect the topological helical edge modes discussed here can be probed in a similar manner. To experimentally measure the topological properties of the Bogoliubov excitations, one can also coherently transfer a small portion of the condensate into an edge mode using Raman transitions \cite{Ernst2009}, and a density wave should be formed along the edge due to an interference with the background condensate \cite{Furukawa2015}.

One straightforward generalization of our work is to consider a AII-class-like excitation band topology of BEC in three dimensions \cite{Fu2007a,Kondo2019a}. Using the language of Krein-space theory developed in this paper, one may also consider various (symmorphic or nonsymmorphic) crystalline-symmetry-protected excitation band topology of weakly interacting BEC in optical lattices in the superfluid phase, in analogous to its fermionic counterpart \cite{Ando2015}. One may also study excitations in the Mott insulator phase, where similar topological structure is expected to occur \cite{Vasic2015,Wu2017}.

Lastly, we note that, despite the well-known effectiveness of the Bogoliubov theory to the weakly interacting bosons, it is interesting to go beyond this approximation and consider higher-order quantum corrections, by using either exact numerical methods or many-body perturbation theory. The fate of PTRS and the associated topological properties could be studied further in the future.

\begin{acknowledgments}
YD acknowledges the support by National Natural Science Foundation of China (under Grant No. 11625522) and the National Key R\&D Program of China (under Grant No. 2016YFA0301604 and No. 2018YFA0306501).
\end{acknowledgments}

\appendix
\section{Properties of two sewing matrices and Pfaffian}
For the sewing matrix $B$ defined in Eq.~\eqref{sewingm}, one can find its explicit matrix elements as follows
\begin{equation}
	\begin{split}\label{smb}
	\quad \krin{u_m(-\vb k),\mathcal T u_n(\vb k)} &= -\krin{u_n(\vb k),\mathcal T u_m(-\vb k)}\\
	&= \krin{u_n(\vb k),B_{ml}(\vb k)u_l(\vb k)}\\
	&= B_{mn}(\vb k).
	\end{split}
\end{equation}
It then follows that $B$ is \textit{unitary} (double indices imply summation)
\begin{equation}
	\begin{split}
	&\quad	B_{mn}(\vb k)B^*_{ln}(\vb k)\\
		 &=\krin{u_m(-\vb k),\mathcal T u_n(\vb k)}\krin{u_l(-\vb k),\mathcal T u_n(\vb k)}^*\\
	&=\krin{u_n(\vb k),\mathcal T u_m(-\vb k)}\krin{\mathcal T u_l(-\vb k),u_n(\vb k)}\\
	&=\bra{\mathcal T u_l(-\vb k)}\Sigma_3 \ketbra{u_n(\vb k)}{u_n(\vb k)}\Sigma_3\ket{\mathcal T u_m(-\vb k)}\\
	&= \krin{\mathcal T u_l(-\vb k), \mathcal T u_m(-\vb k)} \\
	&= \krin{u_m(-\vb k),u_l(-\vb k)}\\
	&=\delta_{ml},
	\end{split}
\end{equation}
and has the property $B_{mn}(\vb k)=-B_{nm}(-\vb k)$ since
\begin{equation}\label{bmnbnm}
	\begin{split}
		B_{mn}(\vb k) &= \krin{u_m(-\vb k),\mathcal T u_n(\vb k)}\\
	&= -\krin{u_n(\vb k),\mathcal T u_m(-\vb k)}\\
	&= -B_{nm}(-\vb k).
	\end{split}
\end{equation}
The sewing matrix for hole bands is related to its particle companion by
\begin{equation}\label{bhole}
	\begin{split}
		[B_{\textnormal{hole}}]_{mn}(\vb k)&=\krin{\Sigma_1 u_m^*(\vb k),\mathcal T \Sigma_1 u_n(-\vb k)^*}\\
	&=\pm \krin{u_m(\vb k),\mathcal T u_n(-\vb k)}^*\\
	&=\pm B_{mn}^*(-\vb k)
	\end{split}
\end{equation}
where $\pm$ corresponds to $P=\tau_3\otimes M$ or $\tau_0\otimes M$ with $\tau_i$ ($\tau_0$), $i=1,2,3$, the standard Pauli (two-by-two identity) matrix and the Hermitian matrix $M$ satisfying $MM^*=-1$. In particular, for the two examples studied in Sec.~\ref{sec3}, the minus sign is picked.

For the sewing matrix $C$ defined in Eq.~\eqref{sewmatcdef}, one can find its explicit matrix elements as follows
\begin{equation}\label{cdefanother}
	\begin{split}
		&\quad \krin{u_m(\vb k),\mathcal P\mathcal T u_n(\vb k)}\\
		&=-\braket{u_m(\vb k)}{\Sigma_3C_{nl}(\mathcal P\mathcal T)^2u_l(\vb k)}\\
		&= C_{nl}\krin{u_m(\vb k),u_l(\vb k)}\\
	&=C_{mn}
	\end{split}
\end{equation}
It then follows that $C$ is \textit{unitary},
\begin{equation}\label{prfcuni}
	\begin{split}
		&\quad C_{mn}(\vb k)C^*_{ln}(\vb k)\\
		&=\krin{u_m(\vb k),\mathcal P\mathcal T u_n(\vb k)}\krin{u_l(\vb k),\mathcal P\mathcal T u_n(\vb k)}^*\\
		&=\krin{\mathcal P u_n(\vb k),\mathcal T u_m(\vb k)}\krin{\mathcal T u_l(\vb k),\mathcal P u_n(\vb k)}\\
		&=\bra{\mathcal T u_l(\vb k)} \Sigma_3\mathcal P \ket{u_n(\vb k)}\bra{u_n(\vb k)}\mathcal P \Sigma_3 \mathcal T \ket{u_m(\vb k)}\\
		&= \bra{\mathcal T u_l(\vb k)} \Sigma_3 \mathcal T \ket{u_m(\vb k)}\\
		&= \krin{u_m(\vb k),u_l(\vb k)}\\
		&=\delta_{ml}
	\end{split}
\end{equation}
\textit{where we used pseudo-unitarity of $\mathcal P$.} And $C$ is \textit{antisymmetric},
\begin{equation}\label{prfcansy}
	\begin{split}
		C_{mn}(\vb k) &=\krin{u_m(\vb k),\mathcal P\mathcal T u_n(\vb k)}\\
		&= -\krin{\mathcal P u_n(\vb k),\mathcal T u_m(\vb k)}\\
		&=-\krin{u_n(\vb k),\mathcal P \mathcal T u_m(\vb k)}\\
		&= -C_{nm}
	\end{split}
\end{equation}
\textit{where we used pseudo-Hermiticity of $\mathcal P$.} One can relate $C$ at $\vb k$ and $-\vb k$ using $B$:
\begin{equation}
	\begin{split}
		&\quad C_{mn}(-\vb k)\\
		 &=\krin{u_m(-\vb k),\mathcal P\mathcal T u_n(-\vb k)}\\
		&= B_{ml}(\vb k)\krin{\mathcal T u_l(\vb k),\mathcal P\mathcal T^2u_{l'}(\vb k)}B_{nl'}(\vb k)\\
		&= B_{ml}(\vb k)\krin{u_{l}(\vb k),\mathcal P\mathcal T u_{l'}(\vb k)}^*B_{nl'}(\vb k)\\
		&= B_{ml}(\vb k)C_{ll'}^*(\vb k)[B^T(\vb k)]_{l'n}
	\end{split}
\end{equation}
which leads to Eq.~\eqref{cbrelation}. Similarly, one can relate the Pfaffian $P$ at $\vb k$ and $-\vb k$ using $B$:
\begin{equation}\label{pfatkmk}
\begin{split}
	P(-\vb k) &= \pf[\krin{u_n(-\vb k),\mathcal T u_m(-\vb k)}]\\
	& = \pf[B_{nl}(\vb k) \krin{\mathcal T u_{l'}(\vb k),u_{l}(\vb k)}B_{ml'}(\vb k)]\\
	& = \pf[B^*_{nl}(\vb k)\krin{u_{l}(\vb k),\mathcal T u_{l'}(\vb k)}B^*_{ml'}(\vb k)]^*\\
	& = \det[B(\vb k)] P^*(\vb k).
\end{split}
\end{equation}
The Pfaffian for the hole bands is also related to its particle companion,
\begin{equation}\label{holepfaffian}
	\begin{split}
		P_{\textnormal{hole}}(\vb k) &=\pf[\krin{\Sigma_1 u^*(-\vb k),\mathcal T \Sigma_1 u_m^*(-\vb k)}]\\
		&=\pm \pf[\krin{u^*(-\vb k),\mathcal T u_m^*(-\vb k)}]\\
		&=\pm \pf[\krin{u(-\vb k),\mathcal T u_m(-\vb k)}]^*\\
		&=\pm P(-\vb k)^*\\
		&=\pm \det[B(\vb k)]^* P(\vb k)
	\end{split}
\end{equation}
where again $\pm$ corresponds to $P=\tau_3\otimes M$ or $\tau_0\otimes M$.

Lastly, consider a gauge transformation only in particle space $\ket{u_n(\vb k)} \rightarrow  R_{nm}(\vb k)\ket{u_m(\vb k)}$, to preserve the orthonormal condition w.r.t. the pseudo inner product, $R$ has to be unitary. It then follows that the Pfaffian becomes
\begin{equation}\label{pdetrp}
\begin{split}
	P(\vb k) &\rightarrow  \pf[R^*_{nl}(\vb k)\krin{u_l(\vb k),\mathcal T u_{l'}(\vb k)} R^*_{ml'}(\vb k)]\\
	& = \det[R^*(\vb k)]P(\vb k).
\end{split}
\end{equation}
The sewing matrix $B$ becomes
\begin{equation}\label{gt4b}
	\begin{split}
		B_{mn}(\vb k)\rightarrow [R^*(-\vb k)]_{ml}B_{ll'}(\vb k)[R^\dagger(\vb k)]_{l'n}.
	\end{split}
\end{equation}
And the sewing matrix $C$ becomes
\begin{equation}
	\begin{split}
		C_{mn}(\vb k)\rightarrow [R^*(\vb k)]_{ml}C_{ll'}(\vb k)[R^\dagger(\vb k)]_{l'n}.
	\end{split}
\end{equation}

\section{Proof of Eq.~\eqref{trp}}
We first relate the Berry connection between two states of the $\lambda$-th pair, using Eq.~\eqref{12x},
\begin{equation}
\begin{split}
	A^{(1)}_\lambda (-k) &= -i\krin{\partial_k \mathcal T u^{(1)}_\lambda (-k),\mathcal T u_\lambda ^{(1)}(-k)}\\
	& = -i\krin{\partial_k u^{(2)}_\lambda  (k),u_\lambda ^{(2)}(k)}+ \partial_k \chi_{k,\lambda }\\
	&= A^{(2)}_\lambda (k)+ \partial_k \chi_{k,\lambda }.
\end{split}
\end{equation}
Then the partial polarization for $l=1$ can be written as
\begin{equation}\label{p1ac}
\begin{split}
	P^{(1)}_\lambda  &= \frac{1}{2\pi}\int_0^\pi \dd{k} \bqty{A^{(1)}_\lambda (k)+A^{(1)}_\lambda (-k)}\\
	&= \frac{1}{2\pi}\bqty{\int_0^\pi \dd{k} A_\lambda (k)+(\chi_{\pi,\lambda }-\chi_{0,\lambda })},
\end{split}
\end{equation}
where $A_\lambda (k)=A_\lambda ^{(1)}(k)+A_\lambda ^{(2)}(k)$ is the full (Abelian) Berry connection. Using the sewing matrix $B$ and the representation Eq.~\eqref{12x}, we have $\frac{\pf[B_\lambda (\pi)]}{\pf[B_\lambda (0)]}=e^{-i \chi_{\pi,\lambda }+i \chi_{0,\lambda }}$, hence Eq.~\eqref{p1ac} becomes
\begin{equation}\label{p1lambda}
	P^{(1)}_\lambda  = \frac{1}{2\pi}\bqty{\int_0^\pi \dd{k} A_\lambda (k)+i \log\pqty{\frac{\pf[B(\pi)]}{\pf[B(0)]}}}.
\end{equation}
Under a $\mathrm{U}(1)$ gauge transformation $\ket{u_\lambda ^{(l)}(k)}\rightarrow e^{i\tilde \chi(k)}\ket{u_\lambda ^{(l)}(k)}$, both terms on the r.h.s. of Eq.~\eqref{p1lambda} induces $2\tilde \chi(\pi)-2\tilde \chi(0)$ with \textit{opposite} sign, thus cancel each other. By writing the full Berry connection as the trace of the $\mathrm{U}(2)$ non-Abelian Berry connection [cf. Eq.~\eqref{nonaa}], $A_\lambda (k)=\Tr \mathcal A_\lambda (k)$, the first term on the r.h.s. of Eq.~\eqref{p1lambda} is manifestly invariant under a $\mathrm{SU}(2)$ gauge transformation $\ket{u_\lambda ^{(l)}(k)}\rightarrow U_{ll'}(k)\ket{u_\lambda ^{(l')}(k)}$; for the second term, due to Eq.~\eqref{gt4b}, the Pfaffian at the PTRIM transforms as $\pf[B(k)]\rightarrow \pf[B(k)]\det[U^*]$ and is also invariant. We conclude that $P_\lambda ^{(1)}$ is $\mathrm{U}(2)$ invariant in analogous to the fermionic case \cite{Fu2006}. Similarly, one can find that $P_\lambda ^{(2)}=\frac{1}{2\pi}\bqty{\int_{-\pi}^0 \dd{k}A_\lambda (k)-i\log\pqty{\frac{\pf[B(\pi)]}{\pf[B(0)]}}}$, hence the symplectic generalization of ``charge" polarization reads
\begin{equation}\label{cpol}
	P_\lambda =P^{(1)}_\lambda +P^{(2)}_\lambda =\frac{1}{2\pi}\int_{-\pi}^\pi\dd{k} A_\lambda (k).
\end{equation}
Using the sewing matrix $B$, we massage Eq.~\eqref{cpol} at $k_2=-\pi$,
\begin{widetext}
\begin{equation}
\begin{split}
		P_\lambda (k_2=-\pi) &=\frac{i}{2\pi}\int_{-\pi}^\pi\dd{k}\sum_{l=1}^2\krin{u^{(l)}_\lambda (k,-\pi),\partial_{k}u^{(l)}_\lambda (k,-\pi)}\\
	& =-\frac{i}{2\pi}\int_{-\pi}^\pi\dd{k}\sum_{l=1}^2\krin{u^{(l)}_\lambda (-k,-\pi),\partial_{k}u_\lambda ^{(l)}(-k,-\pi)}\\
	&=\frac{1}{2\pi}\int_{-\pi}^\pi\dd{k}\Tr[B^*_\lambda (k,\pi)\mathcal A_\lambda (k,\pi) B_\lambda^T(k,\pi)]+\frac{i}{2\pi}\int_{-\pi}^\pi\dd{k}\Tr[B_\lambda ^\dagger(k,\pi)\partial_k B_\lambda (k,\pi)] \\
	&=P_\lambda (k_2=\pi)+\frac{i}{2\pi}\int_{-\pi}^\pi\dd{k}\partial_k\ln \det B_\lambda (k,\pi)\\
	&=P_\lambda (k_2=\pi).
\end{split}
\end{equation}
Hence the \textit{change} of ``charge" polarization under a cycle of $k_2$ from $-\pi$ to $\pi$ vanishes, which is nothing but the fact that Chern number vanishes for a 2D PTR symmetric system. We then consider the symplectic generalization of PTR polarization,
\begin{equation}
	\tilde P_\lambda  =P^{(1)}_\lambda -P^{(2)}_\lambda = \frac{1}{2\pi}\Bqty{\int_0^\pi \dd{k}A_\lambda (k)-\int^0_{-\pi} \dd{k}A_\lambda (k)+2i\log\pqty{\frac{\pf[B(\pi)]}{\pf[B(0)]}}}\label{PTRp1}
\end{equation}
We massage the middle term on the r.h.s. of Eq.~\eqref{PTRp1} using the sewing matrix $B$,
\begin{equation}
\begin{split}
	\int^0_{-\pi} \dd{k}A_\lambda (k) &=-\frac{i\xi_\lambda }{2\pi}\int_0^\pi \dd{k}\sum_{l=1}^2\krin{u_\lambda ^{(l)}(-k),\partial_k u^{(l)}_\lambda (-k)}\\
	&=\frac{1}{2\pi}\int_0^\pi\dd{k} \Tr[B^*_\lambda (k)\mathcal A_\lambda (k) B^T_\lambda (k)]+\frac{i}{2\pi}\int_0^\pi\Tr[B^\dagger(k)\partial_k B(k)]\\
	&= \frac{1}{2\pi}\int_0^\pi\dd{k}A_\lambda (k)-\frac{1}{2\pi i}\int_0^\pi\dd{k}\partial_k\log\det B(k)
\end{split}
\end{equation}
\end{widetext}
Hence Eq.~\eqref{PTRp1} becomes
\begin{equation}
\begin{split}
	&\quad \tilde P_\lambda \\
	& =\frac{1}{\pi i}\bqty{\int_0^\pi\dd{k} \partial_k \log\sqrt{\det[B(k)]}-\log\pqty{\frac{\pf[B(\pi)]}{\pf[B(0)]}}}
\end{split}
\end{equation}
which leads to Eq.~\eqref{trp}.

\section{Proof of Eq.~\eqref{gfusingc}}
We directly massage the definition of Berry connection using the sewing matrix $C$,
\begin{equation}
	\begin{split}
	&\quad	A_\lambda (\vb k)\\
	 &=i \sum_{l=1,2} \krin{u_\lambda ^{(l)}(\vb k),\nabla_{\vb k} u_\lambda ^{(l)}(\vb k)}\\
		&= i\sum_{l=1,2} C_{ll'}(\vb k) \krin{\mathcal P\mathcal T u_{l'}(\vb k),\nabla_{\vb k} \mathcal P\mathcal T u_{l''}(\vb k)} C^*_{ll''}(\vb k)\\
		&\quad + i\sum_{l=1,2} C_{ll'}(\vb k) \krin{\mathcal P\mathcal T u_{l'}(\vb k),\mathcal P\mathcal T u_{l''}(\vb k)} \nabla_{\vb k} C^*_{ll''}(\vb k)\\
		&=-\tr C^* \mathcal A_\lambda ^*(\vb k) C^T-i\tr C^\dagger(\vb k) \nabla_{\vb k} C(\vb k)\\
		&= -A_\lambda (\vb k)-i\tr C^\dagger(\vb k) \nabla_{\vb k} C(\vb k)\nonumber
	\end{split}
\end{equation}
which leads to the first line of Eq.\eqref{gfusingc}. The second line then follows from the identity $\nabla\log\det[U]=\tr[\nabla\log U]=\tr[U^\dagger \nabla U]$, valid for any unitary matrix $U$.
\section{Properties of Wilson loop operator}
Suppose $\ket{w}$ is an eigenvector of the Wilson loop operator $W_{\vb k}$ with the eigenvalue $w$, i.e., $W_{\vb k}\ket{w}=w \ket{w}$. For any $\vb k'$, which satisfies $\vb k'=n \boldsymbol{\delta}_1+\vb k$ (w.l.o.g. $0\leq n\leq N_1$), we have
\begin{equation}
\begin{split}
	&\quad wM^{[\vb k'-N_1\boldsymbol{\delta}_1,\vb k'-(N_1-1)\boldsymbol{\delta}_1]}M^{[\vb k'-(N_1-1)\boldsymbol{\delta}_1,\vb k'-(N_1-2)\boldsymbol{\delta}_1]}\\
	&\quad  \dots M^{[\vb k'-(n+1)\boldsymbol{\delta}_1,\vb k'-n\boldsymbol{\delta}_1]} \ket{w}\\
	 &=M^{[\vb k'-N_1\boldsymbol{\delta}_1,\vb k'-(N_1-1)\boldsymbol{\delta}_1]} M^{[\vb k'-(N_1-1)\boldsymbol{\delta}_1,\vb k'-(N_1-2)\boldsymbol{\delta}_1]}\\
	 &\quad  \dots M^{[\vb k'-(n+1)\boldsymbol{\delta}_1,\vb k'-n\boldsymbol{\delta}_1]} W_{\vb k}\ket{w}\\
	&=W_{\vb k'}M^{(\vb k',\vb k'+\boldsymbol{\delta}_1)}M^{(\vb k'+\boldsymbol{\delta}_1,\vb k'+2\boldsymbol{\delta}_1)} \dots M^{(\vb k-\boldsymbol{\delta}_1,\vb k)}\ket{w}.\nonumber
\end{split}
\end{equation}
Hence $W_{\vb k'}$ also has the same eigenvalue $w$. Namely, \textit{eigenvalues of $W_{\vb k}$ are independent of $k_1$.}

One may write the Wilson loop operator in terms of a non-Hermitian projector,
\begin{equation}
	W_{\vb k}=\Pi_0 \Pi_1 \dots \Pi_{N_1},\nonumber
\end{equation}
where $\Pi_l=\sum_{n\leq n_\tmax} \ketbra{u_n(\vb k+l\boldsymbol{\delta}_1)}{u_n(\vb k+l\boldsymbol{\delta}_1)}\Sigma_3$ is the projector to the occupied subspace, and the entries of $W_{\vb k}$ are given by $[W_{\vb k}]_{mn}=\krin{u_m(\vb k),W_{\vb k}u_n(\vb k)}$. It is therefore manifestly $\mathrm{U}(n_m)$ gauge invariant.

Since eigenvalues of $W_{\vb k}$ are independent of $k_1$, w.l.o.g., we may consider a particular Wilson loop operator
\begin{equation}
	W_{k_2}=\Pi_{-k_1/2,k_2}\Pi_{-k_1/2+\delta_1,k_2} \dots \Pi_{k_1/2,k_2}\nonumber
\end{equation}
where $\Pi_{k_1,k_2}=\sum_{n\leq n_\tmax} \ketbra{u_n(k_1,k_2)}{u_n(k_1,k_2)}\Sigma_3$. Using the sewing matrix $B$, we have
\begin{align}
	&\Pi_{-k_1,-k_2}\nonumber\\
	&=\sum_{n\leq n_\tmax} \ketbra{u_n(-k_1,-k_2)}{u_n(-k_1,-k_2)}\Sigma_3\nonumber\\
	&=\sum_{n,l,l'\leq n_\tmax}B^*_{nl}(\vb k) P \ketbra{u^*_l(k_1,k_2)}{u^*_{l'}(k_1,k_2)}P^\dagger \Sigma_3 B_{nl'}\nonumber\\
	&=P \Pi_{k_1,k_2}^* P\inv \label{kminusk} \\
	&= P\Sigma_3 \Pi_{k_1,k_2}^T \Sigma_3 P\inv,\nonumber
\end{align}
where we used the unitarity of $B$ and the pseudo-unitarity of $P$. It follows that the the Wilson loop at $-k_2$ and at $k_2$ is related,
\begin{equation}\label{WLiden}
\begin{split}
	W_{-k_2} &=\Pi_{-k_1/2,-k_2}\Pi_{-k_1/2+\delta_1,-k_2} \dots \Pi_{k_1/2,-k_2}\\
	&= P \Sigma_3 \Pi_{k_1/2,k_2}^T \Pi_{k_1/2-\delta_1,k_2}^T \dots \Pi_{-k_1/2,k_2}^T \Sigma_3 P\inv\\
	&= P \Sigma_3 W_{k_2}^T \Sigma_3 P\inv .
\end{split}
\end{equation}
Since eigenvalues remain the same under both the transpose and the similarity transformations, \textit{the Wilson loop at $k_y$ and $-k_y$ have the same eigenvalues}.

At $k_2=0$ or $\pi$, the 1D effective Hamiltonian is PTR symmetric, therefore each eigenstate $\ket{\psi}$ has a PTR companion $\mathcal T\ket{\psi}$ with the same energy and are orthogonal w.r.t. the pseudo inner product,
\begin{equation}\label{orth}
	\krin{\psi,\mathcal T \psi}=\braket{\psi}{\Sigma_3 PK\psi}=0.
\end{equation}
These states are also the eigenstates of the Wilson loop $W=W_{0}$ or $W_\pi$. Using Eq.~\eqref{WLiden} and pseudo-unitarity of $P$, we have
\begin{equation*}
	\begin{split}
		w\ket{\psi} &= W\ket{\psi}\\
		&= P\Sigma_3P^\dagger \Sigma_3 W P\Sigma_3P^\dagger \Sigma_3\ket{\psi}\\
		&= P\Sigma_3 W^T P^\dagger \Sigma_3 \ket{\psi}
	\end{split}
\end{equation*}
Multiplying both side from the left by $P^*$, then taking the complex conjugation, we have
\begin{equation}\label{WL2}
	\begin{split}
		w^*PK\ket{\psi} &= \Sigma_3 W^\dagger (P\inv)^\dagger \Sigma_3 K \ket{\psi}\\
		\Leftrightarrow\quad w^* \Sigma_3 PK\ket{\psi} & = W^\dagger \Sigma_3 PK\ket{\psi}
	\end{split}
\end{equation}
where we used $PP^*=-1$ and again pseudo-unitarity of $P$. Since Eq.~\eqref{WL2} shows that the Wilson loop $W$ has a left eigenvector with eigenvalue $w^*$ which is orthogonal to $\ket{\psi}$ due to Eq.~\eqref{orth}, \textit{the right eigenvalue $w$ must be at least twice degenerate at $k_2=0$ or $\pi$}.

Lastly, we note that the Wilson loop operator for the hole bands is related to its particle companion, $W^{\textnormal{hole}}_{\vb k}=\Sigma_1 (W_{-\vb k}^{\textnormal{particle}})^*\Sigma_1=\Sigma_1P^* (W_{\vb k}^{\textnormal{particle}})(P^{*})\inv \Sigma_1$, where in the second equality we used Eq.~\eqref{kminusk}. Thus they have the same Wannier center flow structure. 

\section{Details on mean-field theory and symmetry analysis}
In this appendix, we present a detailed mean-field calculation for two models discussed in the main text. Especially, we consider a more general interaction term with Eq.~\eqref{hint} as a special case. The requirement of the form of interaction in order to get a BdG system with PTRS is examined.
\subsection{The BKM model}
\begin{figure}
	\includegraphics[width=8.4cm]{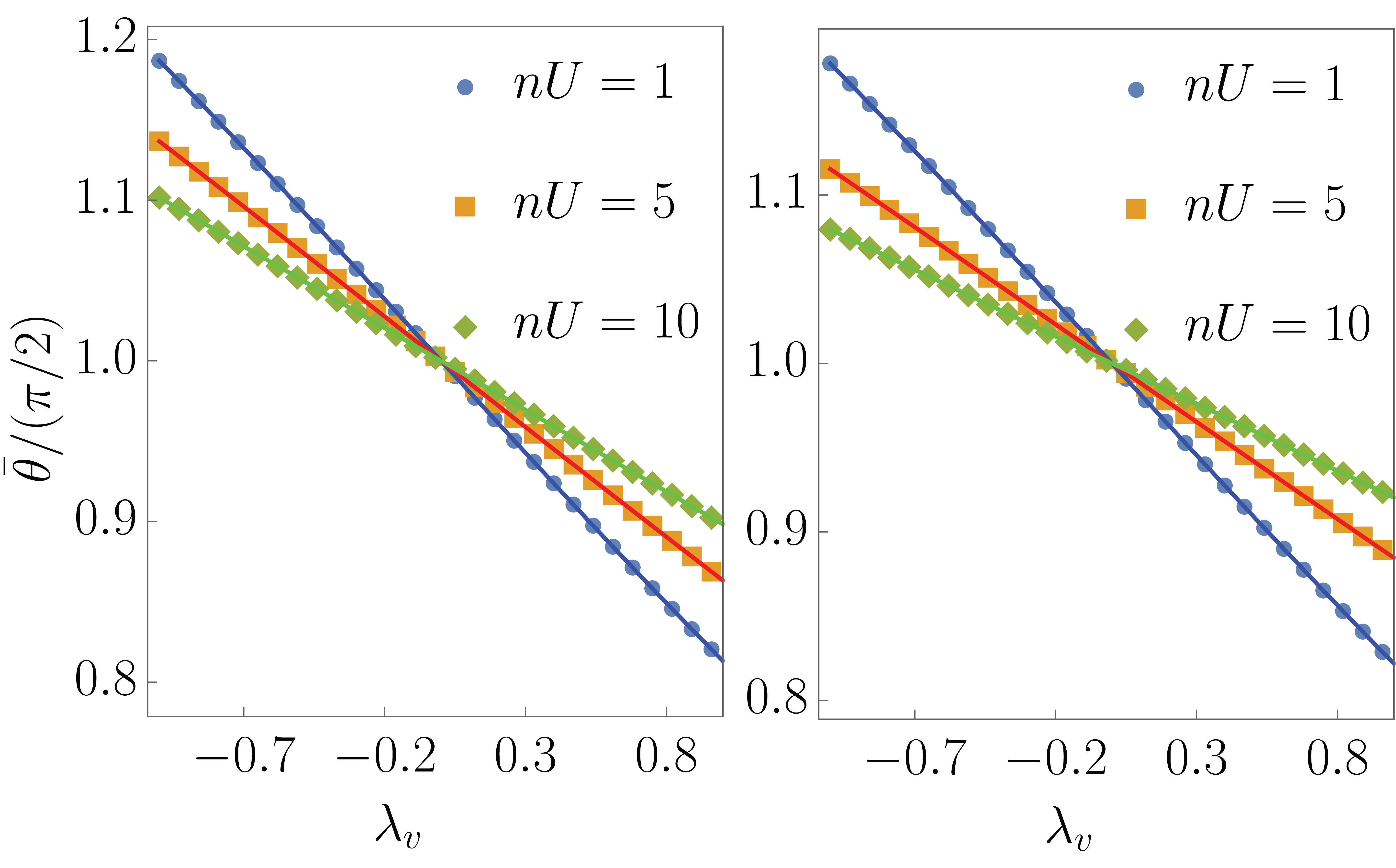}
	\caption{\label{bkmtheta} Mean-field solution of $\theta=2\arctan(\rho_3/\rho_1)$ for the BKM model as a function of sublattice imbalance $\lambda_v$ for interaction anisotropy $\lambda=0.3$ (left) and $\lambda=1.5$ (right), obtained by both minimizing Eq.~\eqref{hmf} numerically using the method of simulated annealing (dots) and minimizing Eq.~\eqref{egp} (left) or Eq.~\eqref{egp1} (right) analytically (solid lines). Note although $\bar\theta$ behaves similarly for $\lambda>1$ and $\lambda<1$, they correspond to different ground state, i.e., Z-ferro and XY-ferro, respectively. Other relevant parameters: $\lambda_s/t=0.06$.}
\end{figure}

We start from the full Hamiltonian, with a generic repulsive interaction, written in momentum space,
\begin{equation}\label{bkmfullh}
\begin{split}
		H&=\sum_{\vb k}a^\dagger_{\vb k}h(\vb k)a_{\vb k}\\
	&\quad +\frac{1}{2M}\sum_{\vb k,\vb p,\vb q,\sigma ss'}U_{ss'} a^\dagger_{\vb k+\vb q,\sigma s}a^\dagger_{\vb p-\vb q,\sigma s'}a_{\vb k,\sigma s'}a_{\vb p,\sigma s},
\end{split}
\end{equation}
where $h(\vb k)$ is given in Eq.~\eqref{bkmh0k}, $M$ is the total number of unit cells, $U_{\uparrow\uparrow}=U_{\downarrow\downarrow}=U>0$, $U_{\uparrow\downarrow}=U_{\downarrow\uparrow}=\lambda U$ and $\lambda>0$ is the interspecies anisotropy. Assuming bosons condense at $\vb \Gamma$, the ground state wave function ansatz is
\begin{equation}
	\ket{\psi}=\frac{1}{\sqrt{\mathcal N!}}\pqty{\sqrt{\mathcal N}\sum_{\sigma s}\psi_{\sigma s}a_{\vb{\Gamma}\sigma s}^\dagger}^{\mathcal N}\ket{0},\nonumber
\end{equation}
where $\mathcal N$ is the total boson number and four complex numbers $\phi_{\sigma s}$ satisfy $\sum_{\sigma s}\abs{\psi_{\sigma s}}^2=1$. Using the following parametrization
\begin{equation}\label{parametrization}
	\pqty{\psi_{A\uparrow},\psi_{A\downarrow},\psi_{B\uparrow},\psi_{B\downarrow}}=\pqty{\rho_1 e^{i \phi_1},\rho_2 e^{i \phi_2},\rho_3 e^{i \phi_3},\rho_4 e^{i \phi_4}},
\end{equation}
the Gross-Pitaevskii (GP) energy functional density becomes
\begin{equation}\label{hmf}
\begin{split}
	\mathcal E_\tgp &= \frac{\mel{\psi}{H}{\psi}}{\mathcal N}\\
	 &= -6t[\cos(\phi_1-\phi_3) \rho_1 \rho_3-\cos(\phi_2-\phi_4) \rho_2 \rho_4]\\
	 &\quad - \lambda_v (\rho_1^2+\rho_2^2-\rho_3^2-\rho_4^2)\\
	&\quad + \frac{nU}{2} - nU(\rho_1^2+\rho_2^2)(\rho_3^2+\rho_4^2)\\
	&\quad - nU(1-\lambda)(\rho_1^2 \rho_2^2 + \rho_3^2 \rho_4^2),
\end{split}
\end{equation}
where $n=\mathcal N/M$ is the particle number density. Its minimization fixes $\phi_1=\phi_3$ and $\phi_2=\phi_4$ (two phases left arbitrary dictated by two U(1) symmetries of the system). For $\lambda < 1$, we expect the XY-ferro state is favored. By setting $\rho_1=\rho_2$ and $\rho_3=\rho_4$, the GP energy functional simplifies to
\begin{equation}\label{egp}
\begin{split}
	\mathcal E_\tgp\vert_{\lambda<1} &=-12t \rho_1\rho_3- 2 \lambda_v (\rho_1^2-\rho_3^2)\\
	&\quad +\frac{nU}{2}[1-4\rho_1^2 \rho_3^2-2(1-\lambda)(\rho_1^4+\rho_3^4)],
\end{split}
\end{equation}
with the constraint $\rho_1^2+\rho_3^2=1/2$. Further introducing $(\rho_1,\rho_3)=(1/\sqrt{2})(\cos \frac{\theta}{2},\sin\frac{\theta}{2})$, then minimizing Eq.~\eqref{egp} finally fixes $\theta=\bar \theta$. For $\lambda > 1$, we expect the Z-ferro state is favored. By setting $\rho_2=\rho_4=0$ (w.l.o.g., assuming $\lambda_v>0$), the GP energy functional simplifies to
\begin{equation}\label{egp1}
	\mathcal E_\tgp\vert_{\lambda>1} = -6t \rho_1 \rho_3 + \lambda_v (\rho_3^2 - \rho_1^2) + nU(\frac{1}{2}-\rho_1^2 \rho_3^2)
\end{equation}
Further introducing $(\rho_1,\rho_3)=(\cos \frac{\theta}{2},\sin\frac{\theta}{2})$, then minimizing Eq.~\eqref{egp1} finally fixes $\theta=\bar \theta $. Above analysis has been confirmed by minimizing Eq.~\eqref{hmf} directly using the method of simulated annealing, as shown in Fig.~\ref{bkmtheta}. The mean-field analysis shows that $\bar \theta$ decreases (increases) from $\pi/2$ when turning on a positive (negative) sublattice potential $\lambda_v$, which physically means that more bosons will condense into $A$ ($B$) sublattice. While the repulsive interaction suppresses this sublattice imbalance, since it favors a uniform configuration. We note $\bar \theta$ is a monotonically decreasing function of $\lambda_v$, but never reaches its extreme values, $0$ or $\pi$, for any finite $\abs{\lambda_v}$.

\begin{figure}
	\includegraphics[width=8.4cm]{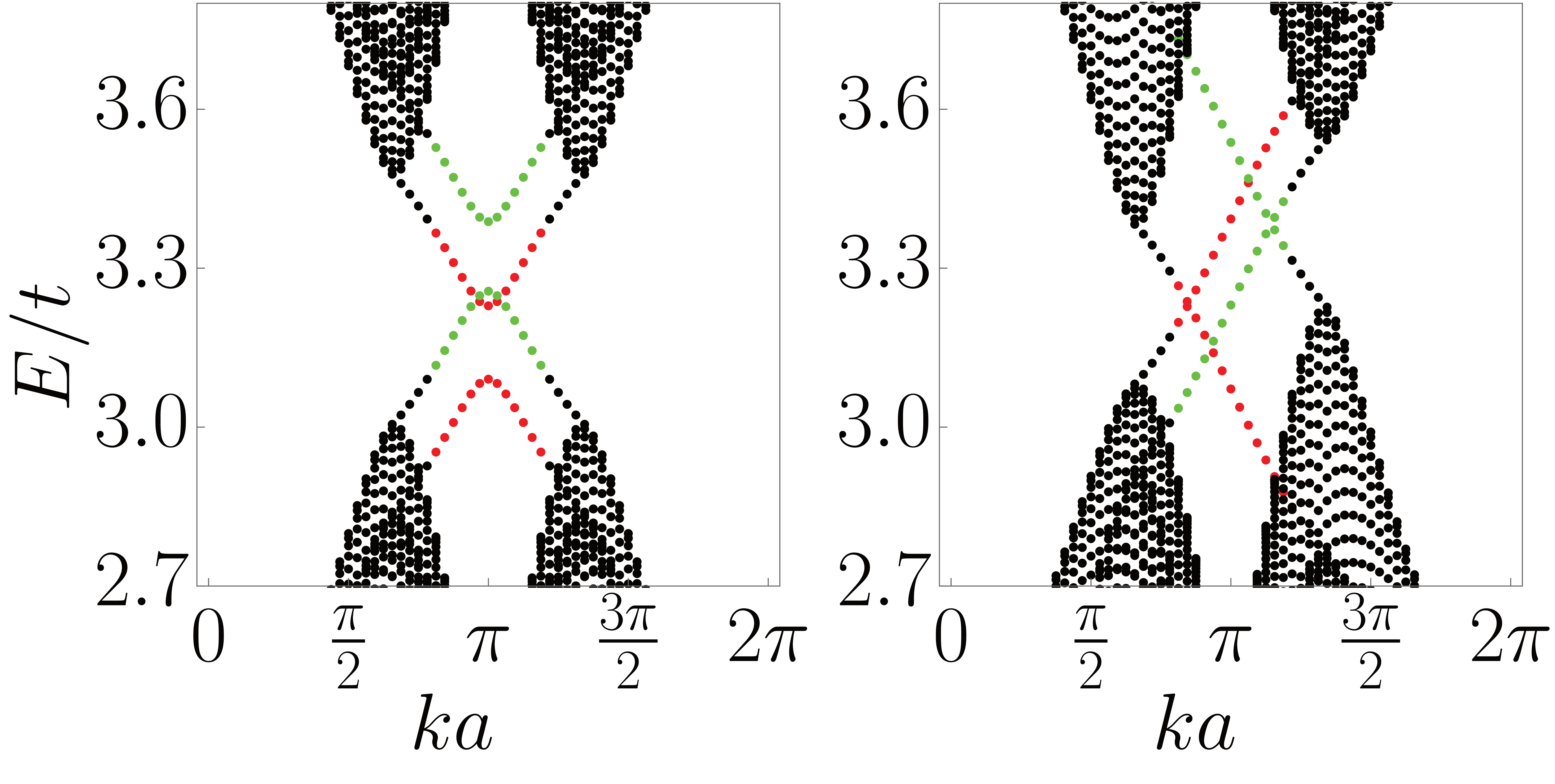}
	\caption{\label{bkmwithoutPTRS} Bogoliubov excitation spectrum near the middle gap of particle bands for BKM model in a strip geometry of $64$ unit cells (each containing $64$ sites) with zigzag edges for $\lambda=0.3$ (left) and $\lambda=1.3$ (right). Red/blue points corresponds to edge modes, whose wavefunctions have more than $80\%$ weight on the leftmost/rightmost unit cell. For both case, the bosonic Kramers' pair is gone. Other relevant parameters: $nU/t=1$, $\lambda_s/t=0.06$ and $\lambda_v/t=0.1$.}
\end{figure}

After obtaining the ground state, we then follow the number-conserving approach \cite{Kawaguchi2012} to the Bogoliubov theory. Making the substitution,
\begin{equation}
	a^{(\dagger)}_{\vb{\Gamma}\sigma s} \rightarrow  \pqty{\mathcal N-\sum_{\vb k\neq\vb \Gamma,\sigma s}a^\dagger_{\vb k,\sigma s}a_{\vb k,\sigma s}}^{1/2}\psi^{(*)}_{\sigma s},\nonumber
\end{equation}
Eq.~\eqref{bkmfullh} can be written, up to the quadratic order in operators, as
\begin{equation}\label{hbog}
	H_{\bog}=\mathcal N\mathcal E_\tgp + \sum_{\vb k\neq \vb \Gamma} a^\dagger_{\vb k}A_{\vb k} a_{\vb k} + (a^\dagger_{\vb k} B a^\dagger_{\vb k}+\hc),
\end{equation}
where
\begin{align}
	A_{\vb k} & = h(\vb k) - \mu I_4 + h_1,\\
	\mu & = \sum_{\sigma s,\sigma' s'} \psi_{\sigma s}^* [h(\vb k)]_{\sigma s,\sigma' s'} \psi_{\sigma' s'},\\
	&\quad + n\sum_{\sigma s s'}U_{ss'}\psi^*_{\sigma s}\psi_{\sigma s'}^*\psi_{\sigma s'}\psi_{\sigma s},\\
	[h_1]_{\sigma s,\sigma' s'} & = n\delta_{\sigma,\sigma'}U_{s s'}(\psi_{\sigma s} \psi^*_{\sigma s'}+\psi^*_{\sigma s'}\psi_{\sigma s'}),
\end{align}
and
\begin{align}
	B_{\sigma s,\sigma' s'} = \frac{n}{2}\delta_{\sigma,\sigma'} U_{ss'} \psi_{\sigma s'}\psi_{\sigma s}.
\end{align}
We plug the mean-field ground-state solution into Eq.~\eqref{hbog}, and rewrite it into a BdG form as discussed in Sec.~\ref{sec11}. For $1>\lambda>0$, the effective Hamiltonian is found to be
\begin{equation}\label{bkmheefl1}
	\begin{split}
	H^\eff_{\vb k}\vert_{\lambda<1} &= H^\eff_{\vb k}\vert_{\lambda=0} + \frac{\lambda nU}{8}\bigg\{i\tau_2\otimes(\cos\bar\theta \Gamma_{13}+\Gamma_{45})\\
	&\quad + \tau_3\otimes \big[\frac{\sin^2\bar\theta}{2} \Gamma_0 + \cos\bar\theta (\Gamma_2 + \Gamma_{13}) +\Gamma_{45}\big]\bigg\},
	\end{split}
\end{equation}
For the PTRS operator defined in Eq.~\eqref{PTRSop1}, the presence of $\Gamma_{13}$ and $\Gamma_{45}$ results in a BdG system \textit{without} PTRS for any $1>\lambda>0$. While, for $\lambda>1$, the effective Hamiltonian turns out to be
\begin{equation}\label{bkmheefg1}
	H^\eff_{\vb k}\vert_{\lambda>1} = H^\eff_{\vb k}\vert_{\lambda=0} + \frac{\lambda n U}{4}\tau_3\otimes \big[ \Gamma_0 + \cos\bar\theta (\Gamma_2 - \Gamma_{15}) - \Gamma_{34} \big].
\end{equation}
due to the presence of $\Gamma_{15}$ and $\Gamma_{34}$, the BdG system does \textit{not} possess PTRS either. In Fig.~\ref{bkmwithoutPTRS}, we show the absence of Bosonic Kramers' pair for both $0<\lambda<1$ and $\lambda>1$. In conclusion, the interspecies interactions breaks the PTRS.

\begin{figure}[t!]
	\includegraphics[width=8.4cm]{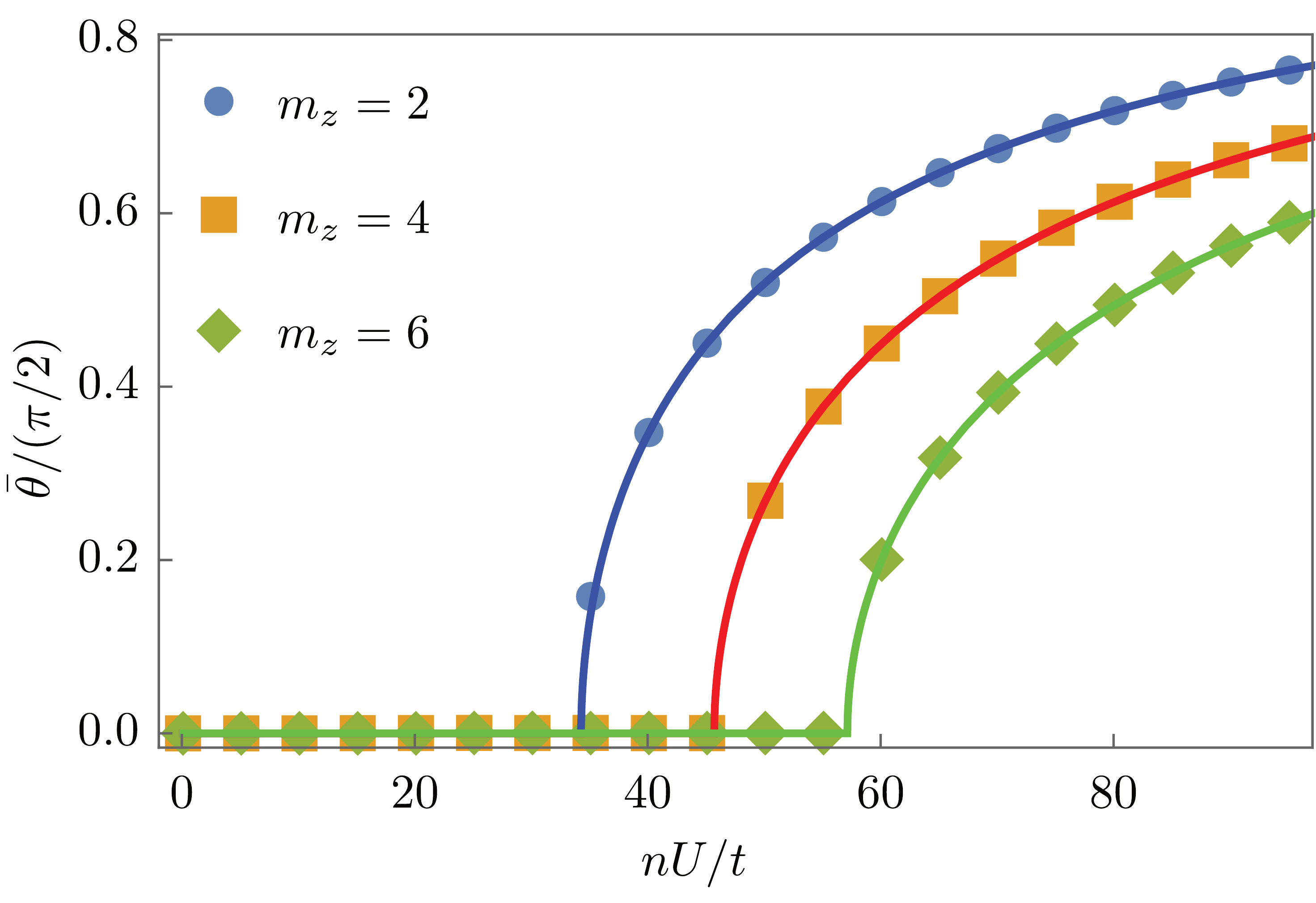}
	\caption{\label{bbhztheta} Mean-field solution of $\theta=2\arctan(\rho_2/\rho_1)$ for the BBHZ model as a function of $nU$ for $\lambda=0.3$, obtained by both minimizing Eq.~\eqref{egpbbhz} numerically using the method of simulated annealing (dots) and minimizing a reduced equation analytically after using the substitution $\rho_1=\rho_3=(1/\sqrt{2})\cos\frac{\bar\theta}{2}$ and $\rho_2=\rho_4=(1/\sqrt{2})\sin\frac{\bar\theta}{2}$ (solid lines). Note for the weakly-interacting region, i.e., for $U/t$ small, we always have $\bar\theta=0$.}
\end{figure}

\begin{figure}[t!]
	\includegraphics[width=8.4cm]{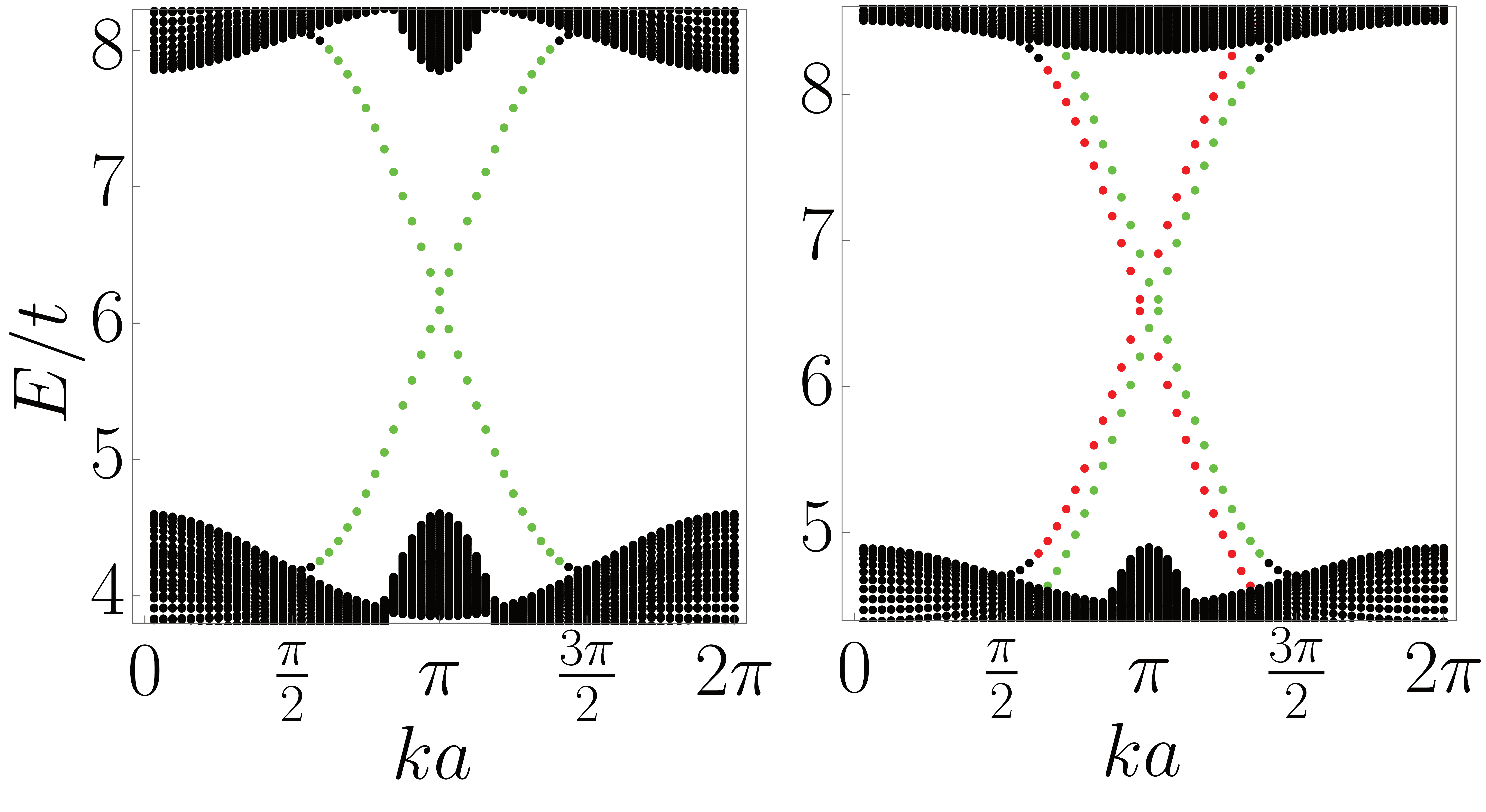}
	\caption{\label{bbhzwithoutPTRS} Bogoliubov excitation spectrum near the middle gap of particle bands for BBHZ model in a strip geometry of $64$ unit cells (each containing 64 sites) for $\lambda=0.3$ (left) and $\lambda=1.3$ (right). Red/blue points corresponds to edge modes, whose wavefunctions have more than $80\%$ weight on the leftmost/rightmost unit cell. Note, for the former case, the left/right edge modes are completely overlapped due to IS. For both case, the bosonic Kramers' pair is gone. Other relevant parameters: $nU/t=t_s/t=1$ and $m_z/t=2.1$.}
\end{figure}

\subsection{The BBHZ model}
The full Hamiltonian, with a generic repulsive interaction, in momentum space reads
\begin{equation}\label{bbhzfullh}
\begin{split}
	&H = \sum_{\vb k}a^\dagger_{\vb k}h(\vb k)a_{\vb k} + \frac{1}{2M}\\
	&\times \sum_{\vb k,\vb p,\vb q,\eta\eta' s s'} U_{\eta s,\eta's'}a^\dagger_{\vb k+\vb q,\eta s}a^\dagger_{\vb p-\vb q,\eta' s'}a_{\vb k,\eta' s'}a_{\vb p,\eta s},
\end{split}
\end{equation}
with $h(\vb k)$ given in Eq.~\eqref{hkbhz}, $U_{\eta s,\eta' s'}=U$ if $\eta=\eta'$ and $s=s'$, $U_{\eta s,\eta' s'}=\lambda U$ otherwise. Physically speaking, we are considering on-site, density-density interaction between all four kinds of bosons (two types $\times$ two pseudospins), this is different from the BKM model, since the latter has two sublattices. Assuming bosons condense at $\vb \Gamma$, which is possible for $m_z$ sufficiently large and positive, the ground state wave function ansatz is
\begin{equation}
	\ket{\psi}=\frac{1}{\sqrt{\mathcal N!}}\pqty{\sqrt{\mathcal N}\sum_{\eta s}\psi_{\eta s}a_{\vb{\Gamma}\eta s}^\dagger}^{\mathcal N}\ket{0},\nonumber
\end{equation}
with four complex numbers satisfying $\sum_{s}\abs{\psi_{\eta s}}^2=1$. Using again the parametrization Eq.~\eqref{parametrization}, the GP energy functional density then becomes \footnote{The GP energy functional is independent of all phase factors $\phi_i$, $i=1,\dots,4$, due to the existence of both two exact U(1) symmetries, and accidental symmetries at mean field level. The latter can be lifted by considering quantum fluctuations, e.g., Ref.~\cite{You2012}.}
\begin{equation}\label{egpbbhz}
	\begin{split}
		\mathcal E_\tgp &= -(4t+m_z)(\rho_1^2-\rho_2^2+\rho_3^2-\rho_4^2)\\
		&\quad + nU\bigg[\frac{1}{2}+(\lambda-1) (\rho_1^2\rho_2^2 +\rho_1^2\rho_3^2\\
		&\quad + \rho_1^2\rho_4^2 + \rho_2^2\rho_3^2 +\rho_2^2\rho_4^2 + \rho_3^2\rho_4^2)\bigg].
	\end{split}
\end{equation}
For $0\leq\lambda<1$, Eq.~\eqref{egpbbhz} is minimized by setting $\rho_1=\rho_3=(1/\sqrt{2})\cos\frac{\bar\theta}{2}$ and $\rho_2=\rho_4=(1/\sqrt{2})\sin\frac{\bar\theta}{2}$ with $\bar\theta$ plotted in Fig.~\ref{bbhztheta}. Note in the region of the weak-coupling limit, i.e., $U/t$ is small, we always have $\bar\theta=0$, which is assumed to be the case in the following discussion. For $\lambda>1$, Eq.~\eqref{egpbbhz} is simply minimized by setting $\rho_1=1$ and $\rho_2=\rho_3=\rho_4=0$ (or exchange $\rho_1$ and $\rho_3$ due to symmetry).

Again, based on the mean-field ground state obtained, we take into account of fluctuations by using the Bogoliubov theory. Making the following substitution in Eq.~\eqref{bbhzfullh},
\begin{equation}
	a^{(\dagger)}_{\vb{\Gamma}\eta s} \rightarrow  \pqty{\mathcal N-\sum_{\vb k\neq\vb \Gamma,\eta s}a^\dagger_{\vb k,\eta s}a_{\vb k,\eta s}}^{1/2}\psi^{(*)}_{\eta s}.\nonumber
\end{equation}
Up to the quadratic order in operators, the Bogoliubov Hamiltonian takes the same form as Eq.~\eqref{hbog}, with $\mathcal E_\tgp$ given in Eq.~\eqref{egpbbhz} and $A,B$ matrices given by
\begin{align}
	A_{\vb k} &= h(\vb k) - \mu I_4 + h_1 \\
	\mu &= \sum_{\eta s,\eta' s'}\psi^*_{\eta s}[h(\vb k)]_{\eta s,\eta' s'}\psi_{\eta' s'}\\
	&\quad +n\sum_{\eta s,\eta' s'} U_{\eta s,\eta' s'} \psi^*_{\eta s} \psi^*_{\eta' s'}\psi_{\eta' s'}\psi_{\eta s} \\
	[h_1]_{\eta s,\eta' s'} &= n  U_{\eta s,\eta' s'}(\psi_{\eta s}\psi^*_{\eta's'}+\psi_{\eta' s'}^*\psi_{\eta' s'})
\end{align}
and
\begin{equation}
	B_{\eta s,\eta' s'}=\frac{n}{2} U_{\eta s,\eta' s'} \psi_{\eta' s'}\psi_{\eta s}.
\end{equation}
Plugging the mean-field ground-state solution into Eq.~\eqref{bbhzfullh}, for $1>\lambda>0$, the effective Hamiltonian is found to be
\begin{equation}
	\begin{split}
		H^\eff_{\vb k} \vert_{\lambda<1} &= H^\eff_{\vb k}\vert_{\lambda=0} + \frac{\lambda nU}{4}   \tau_3\otimes (-\Gamma_{23}+\Gamma_{45}+3\Gamma_0-\Gamma_1)\\
	&\quad +\frac{\lambda nU}{4} i   \tau_2 \otimes (-\Gamma_{23}+\Gamma_{45})
	\end{split}
\end{equation}
Noting that for the PTRS operator defined in Eq.~\eqref{PTRSop2}, we have [for $\mathcal P$ defined in Eq.~\eqref{ISop}, the plus and minus signs on the r.h.s. are exchanged]
\begin{equation}
	\mathcal T [\tau_i \otimes \Gamma_{ab}] \mathcal T\inv = \begin{cases}
		&+\tau_i \otimes\Gamma_{ab} \qfor a=1 \qor b=1,\\
		&-\tau_i \otimes\Gamma_{ab} \qfor a \neq 1 \qand b \neq 1.
	\end{cases}\nonumber
\end{equation}
Hence the presence of $\Gamma_{23}$ and $\Gamma_{45}$ results in a BdG system \textit{without} PTRS (but still has IS) for any $1>\lambda>0$. While, for $\lambda>1$, the effective Hamiltonian turns out to be
\begin{equation}
	H^\eff_{\vb k}\vert_{\lambda>1} = H^\eff_{\vb k}\vert_{\lambda=0}+\frac{\lambda nU}{4}\tau_3\otimes (-\Gamma_{34}+\Gamma_{25}+3\Gamma_0-\Gamma_1)
\end{equation}
Again, due to the presence of $\Gamma_{34}$ and $\Gamma_{25}$, the BdG system does \textit{not} possess PTRS either (but still has IS). In Fig.~\ref{bbhzwithoutPTRS}, we show the absence of bosonic Kramers' pair for both $0<\lambda<1$ and $\lambda>1$. In conclusion, the interspecies interactions again breaks the PTRS.

%


\end{document}